\documentclass[preprint]{elsarticlepp}

\usepackage{lineno,hyperref}  
\modulolinenumbers[1]
\newcommand{\paperOnly}[1]{}

\usepackage[utf8]{inputenc} 
\usepackage{amsmath,amsthm,amssymb,relsize}
\usepackage{float}
\usepackage{extarrows}
\usepackage {tikz} 
\usetikzlibrary{positioning,shapes,fit,calc,backgrounds,matrix,tikzmark}
\usepackage{verbatim}
\usepackage{tkz-graph}
\usepackage{tikz-cd} 
\usepackage{caption}
\usepackage{standalone}
\usepackage{setspace}
\usepackage{subcaption}
\usepackage{mdframed}
\usepackage[resetcount,ruled,boxruled,vlined]{algorithm2e}

\SetCommentSty{mycommfont}
\usepackage{todonotes} 
\usepackage[capitalise]{cleveref}
\usepackage{wrapfig} 

\newcommand{\renamed}{{renamed}}
\newcommand{\renaming}{{renaming}}
\newcommand{\rename}{{rename}}
\newcommand{\ie}{{i.e.}}
\newcommand{\eg}{{e.g.}}
\newcommand{\wrt}{{w.r.t.}}
\newcounter{trs}

\newcommand{\newR}[1]{\refstepcounter{trs}
R_{\mathtt{\thetrs}}\label{#1}}
\newcommand{\newRd}[1]{\refstepcounter{trs}\Rd_{\mathtt{\thetrs}}\label{#1}}

\newcommand{\refR}[1]{R_{\mathtt{\ref*{#1}}}}
\newcommand{\refRd}[1]{\Rd_{\mathtt{\ref*{#1}}}}
\newcommand{\refFd}[1]{\Fd_\mathtt{{\ref*{#1}}}}
\newcommand{\refVd}[1]{\Vd_{\mathtt{\ref*{#1}}}}

\newcommand{\sz}{\ensuremath{\mathsf{sz}}}

\journal{Journal of Logical and Algebraic Methods in Programming}

\newcommand{\N}{\mathbb{N}} 

\newcommand{\Vd}{\mathcal{V}}
\newcommand{\Fd}{\mathcal{F}}
\newcommand{\Rd}{\mathcal{R}}
\newcommand{\Od}{\mathcal{O}}

\newcommand{\T}{\mathcal{T}}

\newcommand{\Pc}{\mathbf{P}}
\newcommand{\NP}{\mathbf{NP}}
\newcommand{\GI}{\mathbf{GI}}

\newcommand{\verti}[1]{{\left\vert\kern-0.25ex\left\vert\kern-0.25ex\left\vert #1 
    \right\vert\kern-0.25ex\right\vert\kern-0.25ex\right\vert}}

\newcommand{\val}[1]{{\left| #1 \right|}}

\newcommand{\sset}[1]{{\left\{ #1\right\}}}
\newcommand{\set}[1]{{\{ #1\}}}

\renewcommand{\subset}{\subseteq}

\DeclareMathOperator{\id}  {id}
\DeclareMathOperator{\add} {add}
\DeclareMathOperator{\suc}{succ}

\DeclareMathOperator{\Var}{Var}
\DeclareMathOperator{\Func}{Func}

\DeclareMathOperator{\ar}{ar}

\DeclareMathOperator{\join}{Join}

\newcommand{\LVE}{\mathbf{LVE}}
\newcommand{\GVE}{\mathbf{GVE}}
\newcommand{\LFE}{\mathbf{LFE}}
\newcommand{\GFE}{\mathbf{GFE}}
\newcommand{\LE} {\mathbf{LE} }
\newcommand{\GE} {\mathbf{GE} }
\newcommand{\SE} {\mathbf{SFE}}
\newcommand{\VSE}{\mathbf{SVE}}

\DeclareMathOperator{\Graph}{Graph}

\newcommand{\lab}{\mathit{lab}}

\DeclareMathOperator{\outdeg}{outdeg}

\renewcommand{\epsilon}{\varepsilon}

\DeclareMathOperator{\Tree}{Tree}

\DeclareMathOperator{\opl}{\sqcup}

\newcommand{\DG}     {\mathbf{DG}}

\newcommand{\OOLDG} {\mathbf{OOLDG}}

\SetKwInOut{Input}{input}
\SetKwInOut{Output}{output}
\SetKwInOut{Runtime}{runtime}

\newtheorem{theo}{Theorem}[section]
\newtheorem{prop} [theo]{Proposition}
\newtheorem{lemma}[theo]{Lemma}
\newtheorem{coro} [theo]{Corollary}

\theoremstyle{definition}
\newtheorem{defi}[theo]{Definition}
\newtheorem{ex}  [theo]{Example}
\newtheorem{remark}[theo]{Remark}

\usepackage{extarrows}
\usepackage{environ}
\NewEnviron{killcontents}{}

\begin{document}

\begin{frontmatter}

\title{Complexity of Deciding Syntactic Equivalence up to Renaming for Term Rewriting Systems\\ (Extended Version)}

\author[CAM]{Michael Christian Fink Amores\corref{mycorrespondingauthor}}
\cortext[mycorrespondingauthor]{Corresponding author}

\ead{mcf61@cam.ac.uk}
\address[CAM]{University of Cambridge, UK}

\author{David Sabel}
\ead{dsabel@online.de} 

\begin{abstract}Motivated by questions from program transformations,
eight notions of isomorphisms between term rewriting systems are defined, analysed, and classified. 
The notions include global isomorphisms, where the renaming 
of variables and function symbols is the same for all term rewriting rules of the system, 
local ones, where a single renaming for every rule is used, and
a combination, where one symbol set is renamed globally while the other set is renamed locally. 
Preservation of semantic properties like convertibility and termination is analysed for the different isomorphism notions.
The notions of templates and maximal normal forms of term rewriting systems are introduced and algorithms to efficiently compute them are presented.
Equipped with these techniques, the complexity of the underlying decision problems of the isomorphisms are analysed and either shown
to be efficiently solvable or proved to be complete for the graph isomorphism complexity class.
\end{abstract}


\end{frontmatter}


\section{Introduction}
\paragraph*{Motivation and Goals}
We are interested in program transformation and optimisation. Given a programming language (or a representation of it, like a core language) an important  question is if two programs are equal. 
While there are several possibilities of defining semantic equality of programs, 
syntactic equality is often an easy definition and very often used implicitly where sometimes it means the same objects, but often some renamings are allowed (like renaming of function and procedure names, renaming of formal parameters, $\alpha$-renaming for languages with binders). In deduction and inference procedures on programs and in compilers where program transformations and optimisations are applied to programs, (syntactic) equality has to be decided, and hence we want to investigate variations and the complexity of 
syntactic equivalences.
We consider term rewriting systems (TRSs, for short) since they are a well-studied formalism to represent computations and programs (\eg,~they are used as a target language for showing termination of real programs \cite{Brockschmidt-et-al:2021,Giesl-et-al:2011}), 
including non-deterministic programs.
Also, other applications like deduction systems use TRSs in their core and equality also matters for those systems.
Thus, in this paper, we consider the question of whether two TRSs 
are syntactically equivalent, where we allow renaming of variable names and/or renaming of function names.
However, an appropriate
definition of syntactic equality depends on the concrete application.
In our view, fundamental requirements of syntactic equality are that it is reflexive, transitive, and symmetric (and thus  an equivalence relation), and that it is decidable only by inspecting the syntax of the systems but not the semantics, thus we do not want to execute the TRSs to decide syntactic equality.

We discuss some example applications of syntactic equivalence.
An example from program transformation is deciding the applicability of
``common subexpression elimination'', which identifies duplicated ({\ie},~equal) code and shares it. 
Another application is to identify equal programs in example program databases that are used for tests, benchmarks, and competitions. 
Having duplicated problems in the problem set (without knowing it) can distort benchmark and competition results. 
For TRSs, an application is Knuth-Bendix-completion where new rewriting rules are added during the completion procedure.
Here one has to check whether the new rules are new or whether they are already present in the system (perhaps with variables renamed, but in this application, renaming of function symbols is not requested).

To cover the different notions of syntactic equivalence, we
use a systematic approach and thus 
define different notions of ``syntactic equivalence up to renaming'' for 
TRSs, we classify them, and we analyse the
complexity of their corresponding decision problem. 
In TRSs two symbol sets occur: variables, where renaming corresponds to renaming of formal parameters, and function symbols, where renaming corresponds to renaming names of functions or data. Our systematic analysis is then based on three ``levels'' of renaming:
%
renamings can be done locally (there are renamings for each rule of a TRS), 
globally (there is a common renaming for all rules of a TRS)
or not at all (renaming of neither symbol set is allowed).

\paragraph*{Contributions and Results}
As main contribution of this paper, we first introduce eight notions of equivalence by considering syntactic-isomorphisms between TRSs, so-called TRS isomorphisms, and then analyse the relationship between these notions, resulting in a hierarchy of equivalence notions (\cref{prop:implications}).
We distinguish local equivalence (local renamings of variables, functions, or both), standard equivalence (local renaming of one set (functions/variables) and global renaming of the other set (variables/functions)) and global equivalence (global renaming of variables, functions, or both). To ensure that all notions are equivalence relations, we have to require for local and standard renamings, that the TRSs are in a \emph{normal form}, which means that there exists no local renaming (of variables, or functions, resp.) that makes two rules of the TRS
equal. 

Concerning complexity,
we show that the (three) local equivalence notions can be solved in polynomial time (\cref{theo:localisoinp}) and thus their decision problems are in $\Pc$.
The other five (global and standard) equivalence notions are shown to be $\GI$-complete (\cref{theo:TRSisos}), where $\GI$ is the complexity class corresponding to the graph isomorphism problem.

We discuss the usability of the equivalences. The most common equivalences seem to be:
\begin{enumerate}
 \item 
Renaming the variables only locally, since often a given signature is fixed and thus TRSs are considered as syntactic equal w.r.t.~this fixed signature, while locally renaming of variables is permitted: this follows the principle ``names of formal parameters do not matter'', \ie, whether a TRS has a rule  $f(x) \to x$  or $f(y) \to y$ should make no difference, regardless of whether variables $x$ and $y$ appear in other rules.
\item
If the signature is not fixed, renaming of functions is allowed.
Then one usually wants to have a global renaming of functions (since otherwise and as we show in \cref{prop:semantic:f-local}, semantic equivalences like conversion equality are not preserved by the equivalence).
This is represented by one of our standard equivalences.
\item Disallowing local renamings (of variables) but permitting global renaming of functions  has the nice property that the corresponding equivalences are equivalence relations without requiring the TRSs to be in normal form. Thus, it allows equating TRSs that contain ``variants'' of rules (e.g. $f(x,y) \to g(a,x)$ and $f(y,z) \to g(a,y)$). This equivalence is represented by one of our global equivalences.
\end{enumerate}
Semantic properties of the syntactic equivalences are analysed in \cref{subsec:semantic}, where global renaming of functions and global or local renaming of variables is shown to be compatible with rewriting, convertibility and termination (see \cref{theo:semantic:compatibility}) while local renaming of functions may be incompatible. Thus, latter equivalences seem to be counter-intuitive and there seem to be no obvious applications that use such equivalences of TRSs. 
However, we see at least two reasons to include them in our study. 
The first reason is a complexity-theoretic one. 
Since standard and global equivalences are $\GI$-complete, it is interesting to look for related problems that are in $\Pc$ and may give new insights into the open question of the complexity class $\GI$. 
The second reason is that for showing global or standard non-isomorphism of TRSs, it is a fast (but incomplete) procedure to refute local equivalence (in polynomial time), since, as we show in \cref{prop:implications}, this implies (global and standard) non-isomorphism of the TRSs.
Due to our polynomial encoding between graphs and TRSs this test could also be used to refute graph isomorphism. 

The eight equivalences only form a core set of equalities, since other equivalence notions can be built from them. For instance, one can define equivalences that first add function symbols to the signatures of both TRSs such that both TRSs have enough function symbols to apply one of our equivalences. Another example of an equivalence variant is to first compute the normal forms (w.r.t.~function symbols or variables) of both TRSs and thereafter compare them with one of our equivalences that require a normal form.
We discuss some variations and generalisations in \cref{subsec:compatibility}.

\paragraph{Related work}
Several isomorphism problems in this paper are shown to be $\GI$-complete. 
For the graph isomorphism problem as a complexity class $\GI$,
it is known that $\Pc \subseteq \GI \subseteq \NP$ holds,
but it is unknown for both $\subseteq$-relations whether they are strict.
Proving $\GI$-completeness of a problem indicates the hardness of the problem and that there is currently no known polynomial-time algorithm for it.  Details on the complexity class $\GI$ can, for instance, be found in \cite{Schoening:88,Kobler-et-al:92}.
A recent overview from a theoretical and a practical perspective on the graph isomorphism problem is given in \cite{grohe-schweitzer:2020}. Several related problems are shown to be $\GI$-complete in \cite{booth-colbourn:77,zemlyachenko-et-al:85}.

Several works consider equivalence of TRSs, while most (for instance \cite{Toyama86,plaisted:93,Terese03})
consider conversion equivalence 
$\xleftrightarrow{*}$ which coincides with the equivalence of the equational theory induced by the TRS, as shown by Birkhoff's Theorem. However, this equivalence is  semantic and not only syntactic.
This also holds for normalization equivalence, which, for instance, is investigated in \cite{Hirokawa-Middeldorp-Sternagel:14} (based on the work in \cite{Metivier83}) and holds if the normal forms of TRSs are the same. 
A commonality with our work is that \cite{Hirokawa-Middeldorp-Sternagel:14} considers variant-free TRSs (which we call TRSs in $\Vd$-normal form) where a variant of a TRS rule is the rule with renamed variables. 
A notion of syntactic equivalence of TRSs is considered in \cite{Mohan90}, based on subsumption of rewriting rules where a rule is subsumed by another rule if it is an instance of this rule. 
However, complexity of the equivalence is not analysed. 
Most of the works use an (arbitrary) but fixed signature and therefore do not include the renaming of function symbols.

A related problem to equality of TRSs is equality of terms and expressions in
(formal)
languages.
In \cite{basin:1994} term equality including associative-commutative functions 
and commutative binding operators is shown to be $\GI$-complete, while the problem 
is in $\Pc$ without binding operators. Equality of TRSs is not considered.
However, one may use the result to show that equivalence of TRS is in $\GI$ by encoding the given TRSs into single terms with  
associative-commutative operators. 

For $\GI$-hardness on the other hand, an encoding in the opposite direction would be required, which seems to be non-trivial.
Thus, we settled on a more elementary approach to prove our $\GI$-completeness results.

In \cite{schmidt-schauss-et-al:2013} the complexity of $\alpha$-equivalence in lambda-calculi with letrec-bindings was analysed and shown to be $\GI$-complete. 
Although our first-order syntax can be viewed as a sublanguage of these higher-order expressions, we cannot reuse any part of the $\GI$-completeness proof:
for proving, that our isomorphism problem on TRSs is in $\GI$, the work in \cite{schmidt-schauss-et-al:2013} does not include rewriting rules and TRSs
(only isomorphism of plain letrec-expressions is considered).
For proving $\GI$-hardness, no reuse of \cite{schmidt-schauss-et-al:2013} is possible because the first-order-syntax is a sublanguage of the higher-order-language, which makes hardness-proofs in general a stronger result (since the problem in the smaller language is shown to be hard) and requires a new proof. While \cite{schmidt-schauss-et-al:2013} encodes graphs a letrec-expressions, we encode graphs as TRSs. A commonality is that we use the formalism of outgoing-ordered labelled directed graphs and their isomorphism problem, which was shown to be $\GI$-complete in \cite{schmidt-schauss-et-al:2013}. 
Finally, compared to this work we consider and classify different notions of syntactic equivalence.

Our proof method for showing that equalities, based on so-called local renamings,  
 is known as the template-method which was used before in \cite{rosenkrantz-hunt:85}, where certain structural homomorphisms for context-free (and regular) grammars were analysed and shown to be in $\Pc$ or $\GI$- or $\NP$-complete.

\paragraph{Outline}
In \cref{sect:prelim} we define necessary notions and notation.
In \cref{sect:trs} we introduce various types of isomorphisms on TRSs and discuss their semantic properties.
In \cref{sect:templates} we first define templates and maximal normal forms of TRS and then show that they can be computed efficiently. 
Equipped with these tools, we show that the three local TRS isomorphisms are in $\Pc$.
In \cref{globalisos} we show that the reamining five TRS isomorphisms are complete for the graph isomorphism complexity class.
We conclude in \cref{sect:concl}.

\section{Preliminaries on Term Rewriting Systems\label{sect:prelim}}
In this section, we recall the required 
terminology
on term rewriting (see also e.g.~\cite{baader-nipkow:98,ohlebusch:2002,Terese03}).

We define terms as follows, where we -- unlike other definitions -- allow the set of variables to be finite. This is helpful when treating the concrete terms that (syntactically) appear as left- and right-hand sides in term rewriting rules.
Let $\Fd$ be a finite set of \emph{function symbols}, where each $f \in \Fd$ has a fixed arity $\ar(f) \in \N_0$. 
Let $\Vd$ be a disjoint, finite or countably infinite
set of \emph{variables}. \emph{Terms}  $T(\Fd,\Vd,\ar)$ (over function symbols $\Fd$ and variables $\Vd$) 
are defined inductively: 
$x \in \Vd$ is a term, any \emph{constant} (\ie~$f \in \Fd$, with $\ar(f)$ = 0) is a term, and 
if $f\in \Fd$, $\ar(f)\ge 1$, $t_1,\ldots,t_{\ar(f)} \in T(\Fd,\Vd,\ar)$ then $f(t_1,\ldots,t_{\ar(f)}) \in T(\Fd,\Vd,\ar)$. 
The \emph{length} $|t|$ of a term $t\in T(\Fd,\Vd,\ar)$ is its total number of variable and function symbol appearances.
For convenience, we sometimes omit $\ar$ and  write $T(\Fd,\Vd)$.

We use letters $c,d,\ldots$ 
for constants, $f,g,h,\ldots$
for arbitrary functions symbols, and 
$x,y,z$ for variables.
For a term $t$, $\Var(t)$ denotes the variables and $\Func(t)$ the function symbols occurring in $t$.
For convenience, we assume that the set of variables in a term rewriting system is finite. This will ease notation when we define mappings between them.
A \emph{term rewriting system (TRS)} over terms $T(\Fd,\Vd)$ (where $\Vd$ is finite) is a finite set $\Rd$ of rules $\Rd = \{\ell_1 \to r_1,\ldots,\ell_n \to r_n\}$, where $\ell_i,r_i \in T(\Fd,\Vd)$ are terms, $\ell_i$ is not a variable, and $\Var(r_i) \subseteq \Var(\ell_i)$. 
  The \emph{length} $|\ell r|=|\ell|+|r|$ of a rewriting rule $\ell\to r$ is the length of the concatenation of terms $\ell$ and $r$.

We write $R=(\Fd,\Vd,\Rd)$ for a TRS to make the function symbols and the variables explicit.
Since in practice TRSs are often characterised by their rewriting rules only and the underlying symbol sets are determined implicitly or as part of the initial design, assume w.l.o.g. symbol sets to be minimal, \ie~$\Vd=\bigcup_{\ell\to r\in \Rd}\Var(\ell)$ and $\Fd=\bigcup_{\ell\to r\in \Rd}\Func(\ell)\cup \Func(r)$.
We will later (to an extend)
 relax this restriction when talking about isomorphisms.

We define the rewrite relation of a TRS as usual: 
let $C$ denote a \emph{context}, \ie~a term with a hole $[\cdot]$ and let
$C[t]$ be the context $C$ where the hole is replaced by term $t$.
For a TRS $R$ with rule set $\Rd$, the rewrite relation $\to_R$ is defined by
$C[\sigma(\ell)] \to_R C[\sigma(r)]$  for any $C$, any rule $\ell \to r \in \Rd$,
and $\sigma$ a substitution that replaces finitely many variables by terms.

With $\xrightarrow{*}_R$ we denote the reflexive-transitive closure of $\to_R$, \emph{convertibility}
$\xleftrightarrow{*}_{R}$ is the reflexive-transitive closure of the symmetric closure $\xleftrightarrow{}_R$ of $\to_R$.
A TRS $R$ is \emph{terminating} if there is no infinite sequence of $\to_R$-steps.

For a function $f : A \to B$, its \emph{extension} to set $X \subseteq A$ is $f(X) = \{f(x) : x \in X\}$. 
We denote the \emph{restriction} of $f$ to $X$ with $f|_X$.
With $\set{a_1\mapsto b_1,\ldots,a_n\mapsto b_n}$ we denote the map $f: \set{a_1,\ldots,a_n}\to \set{b_1,\ldots,b_n}$, mapping $a_i$ onto $b_i$.
For disjoint sets $A$ and $B$ and maps $f:A\to C$ and $g: B\to D$, let $f\opl g: A\cup B\to C\cup D$ be the unique map with  $(f\opl g)|_A=f$ and $(f\opl g)|_B=g$. This can be generalised to arbitrary many, pairwise disjoint sets.
\begin{ex}
\label{ex:TRS}
Let $\Fd=\set{f,h,g,c}$, $\Vd=\{x,y\}$ and $\ar=\{c\mapsto 0, g\mapsto 1, h\mapsto 1, f\mapsto 2\}$.
Rewriting rules over $T(\Fd,\Vd,\ar)$ include 
$$\text{$h(y)\to f(c,y)$, $f(g(x),y)\to g(x)$, $h(g(x))\to g(x)$ or $h(y)\to c$.}$$
Recall the three conditions on TRSs: (1) \emph{Left and right side of a rule have to be valid terms}, (2) \emph{the left side is not allowed to be a single variable}, (3) \emph{the right side does not introduce new variables}.
Then $f(x)\to c$ violates (1), $x\to h(x)$ violates (2) and $g(x)\to h(y)$ violates (3).
\end{ex}

To model the concept of rule-remapping,
we introduce term homomorphisms, which allow us to modify underlying term sets.

\begin{defi}[Term Homomorphisms and Term Isomorphisms]
\label{defi:termhom}
A \emph{term homomorphism} from term set $T(\Fd,\Vd,\ar)$ into $T(\Fd',\Vd',\ar')$ 
is a pair of maps $\phi=(\phi_{\Fd},\phi_{\Vd}): 
(\Fd \to \Fd',\Vd \to \Vd')$ 
such that 
 $\ar(f)=\ar'(\phi_{\Fd}(f))$ for every $f\in \Fd$, canonically extended to all terms $T(\Fd,\Vd,\ar)$ via $\phi(x)=\phi_{\Vd}(x)$ for $x\in \Vd$ and 
$\phi(f(t_1,\ldots,t_{\ar(f)}))=\phi_{\Fd}(f)(\phi(t_1),\ldots,\phi(t_{\ar(f)}))$ for any $f\in \Fd$, terms $t_1,\ldots,t_{\ar(f)}\in T(\Fd,\Vd,\ar)$.
\emph{Concatenation} of term homomorphisms $\phi,\phi'$ is defined entry-wise, i.e. $\phi\circ \phi' = (\phi_\Fd,\phi_\Vd)\circ(\phi'_\Fd,\phi'_\Vd)=(\phi_\Fd\circ\phi_\Fd',\phi_\Vd\circ\phi_\Vd')$.
If $\phi$ is bijective, it is called a \emph{term isomorphism} and its \emph{inverse} is simply denoted by $\phi^{-1}=(\phi_{\Fd}^{-1},\phi_{\Vd}^{-1})$.
Moreover,  $\phi$ is called 
\emph{$\Vd$-(/$\Fd$)-invariant}, if $\Vd=\Vd'$ ($\Fd=\Fd'$) and $\phi_{\Vd}=\id_{\Vd}$ ($\phi_{\Fd}=\id_{\Fd}$).

\end{defi}
For all intense and purposes, we allow ourselves to abuse notation and treat a term homomorphism $\phi$ as a single map $\phi:(\Fd\cup \Vd)\to (\Fd'\cup \Vd')$ where additionally $\phi(\Vd)\subseteq \Vd'$ and $\phi(\Fd)\subseteq \Fd'$.
In this sense, $\phi_{\Vd}$ can be seen as the restriction
$\phi|_{\Vd}$ (the same for $\phi_{\Fd}$). We sometimes use the restriction notation.


\begin{coro}
\label{coro:termisos}
By \cref{defi:termhom}, inverses and compositions of term isomorphism are again term isomorphisms.
\end{coro}

Now we can specify certain normal forms of TRSs by prohibiting ``equivalent'' rewriting rules, based on manipulation by term isomorphisms.

\begin{defi}[Equivalence of Rewriting Rules, Normal Forms]
\label{defi:normal form}
Let $R=(\Fd,\Vd,\Rd)$ be a TRS.
For two distinct rewriting rules
$\ell\to r,\ell'\to r'\in \Rd$, we say that 
$\ell\to r$ and $\ell'\to r'$ are \emph{($\Vd$-/$\Fd$-)equivalent}, iff
there is a ($\Fd$-/$\Vd$-invariant) term isomorphism $\phi: (\Fd\cup\Vd)\to (\Fd\cup\Vd)$ with $\phi(\ell)\to \phi(r)=\ell'\to r'$. If $\Rd$ does not contain any pair of ($\Vd$-/$\Fd$-)equivalent rewriting rules, then we say that $R$ is in \emph{($\Vd$-/$\Fd$-)normal form}.

\end{defi}
Clearly, a TRS in normal form is already in
$\Vd$-/$\Fd$-normal form.

\begin{ex}[Cont.\,of~\cref{ex:TRS}]
\label{ex:TRSHom}
Both 
$\phi_1 =\{x\mapsto y,y\mapsto x,c\mapsto c, g \mapsto g, h\mapsto h,f\mapsto f\}$
\text{and }
$\phi_2  =\{x\mapsto x,y\mapsto y,c\mapsto c, h \mapsto g, g\mapsto h,f\mapsto f\}$
  are valid term isomorphisms on $T(\Fd,\Vd)$ itself, with the former being $\Fd$-invariant and the latter $\Vd$-invariant.
Moreover, for 
$$\newRd{trs-1}=\{f(g(x),y)\to h(x),f(h(x),y)\to g(x),f(g(y),x)\to h(y)\},$$
$(\Fd,\Vd,\refRd{trs-1})$ is neither in $\Vd$-normal form:
$$\phi_1(f(g(x),y))\to \phi_1(h(x))=f(g(y),x)\to h(y),$$ 
since $f(g(x),y)\to h(x)$ and $f(g(y),x)\to h(y)$ are $\Vd$-equivalent, 
nor in $\Fd$-normal form:
$$\phi_2(f(g(x),y))\to\phi_2(h(x))=f(h(x),y)\to g(x),$$ 
since $f(g(x),y)\to h(x)$ and $f(h(x),y)\to g(x)$ are $\Fd$-equivalent.
\end{ex}

\section{A Hierarchy of TRS Isomorphisms\label{sect:trs}}
We define 
notions of isomorphism between TRSs together with equivalences induced by different term isomorphisms. Afterwards we analyse their relationship with semantic properties of TRSs.

\subsection{Classification of TRS Isomorphisms}

Variables and function symbols can each be {\renamed} locally, globally, or not at all, resulting in a total of eight non-trivial, distinct isomorphism types, after combining all possibilities.
Global {\renaming} requires a common symbol-mapping applied to the whole rule set, while local {\renaming} allows separate symbol-mappings for each rule in the set.
For so-called \emph{global TRS isomorphisms}, we will define variants that require the term isomorphisms to be $\Fd$- or $\Vd$-invariant, resp..
For the local ones, we define variants 
(so-called \emph{standard TRS isomorphisms}), where all local mappings have something in common, \eg,~mappings that have a common renaming on the variables, or mappings that have a common renaming on the function symbols. 

\begin{table}[t]
\centering
\begin{tabular}{|@{\,}c@{\,}|c@{\,}|@{\,}c@{\,}|@{\,}c@{\,}|@{\,}c@{\,}|@{\,}c@{\,}|@{\,}c@{\,}|c@{\,}|}
\hline
 $\cong$ & $\Vd$-inv. & $\Vd$-global & $\Vd$-local 
& $\Fd$-inv. & $\Fd$-global & $\Fd$-local & Requ.\,normal form\\
\hline
\multicolumn{8}{|@{}c@{}|}{\fcolorbox{black!10!white}{black!10!white}{\parbox{.95\textwidth}{\centering Global TRS Isomorphisms}}}
\\
\hline
$\GE$  & & \checkmark & & & \checkmark &  & --\\
$\GVE$ & & \checkmark & & \checkmark & &  & --\\
$\GFE$ & \checkmark  & & & & \checkmark & & --\\
\hline
\multicolumn{8}{|@{}c@{}|}{\fcolorbox{black!10!white}{black!10!white}{\parbox{.95\textwidth}{\centering Standard TRS Isomorphisms}}}
\\
\hline
$\VSE$ & & \checkmark  & & & & \checkmark & $\Fd$-normal form\\
$\SE$  & & & \checkmark  & & \checkmark & & $\Vd$-normal form\\
\hline
\multicolumn{8}{|@{}c@{}|}{\fcolorbox{black!10!white}{black!10!white}{\parbox{.95\textwidth}{\centering Local TRS Isomorphisms}}}
\\
\hline
$\LE$  & & & \checkmark  & & & \checkmark & normal form\\
$\LVE$ & & & \checkmark & \checkmark  & & & $\Vd$-normal form\\
$\LFE$ & \checkmark & & & & & \checkmark & $\Fd$-normal form\\
\hline
\end{tabular}
\caption{Categorisation and comparison of TRS isomorphisms}
\label{tab:TRSisos}
\end{table}

\footnotetext[1]{
Emphasising that we define 
\underline{e}quivalences,
the labels of all relations end with letter ``E'' (see \cref{lem:equrel}).}
\begin{defi}[TRS Isomorphisms{\footnotemark[1]{}}]
\label{defi:TRSiso}
Let $R=(\Fd,\Vd,\Rd)$ and $R=(\Fd',\Vd',\Rd')$
be TRSs.
\begin{enumerate}
 \item TRSs $R$ and $R'$ are \emph{globally isomorphic}, written $R\cong_{\GE}R'$, if there exists a term isomorphism $\phi:(\Fd\cup \Vd)\to (\Fd'\cup \Vd')$ with $\phi(\Rd):=\{\phi(\ell)\to \phi(r): \ell\to r\in \Rd\}=\Rd'$. Map $\phi$ is then called a \emph{global isomorphism}.
 \item TRSs $R$ and $R'$ are \emph{$\Vd$-globally isomorphic}, written $R\cong_{\GVE}R'$,
 if there exists a global isomorphism $\phi$ that is $\Fd$-invariant.
 Map $\phi$ is then called a \emph{$\Vd$-global isomorphism}.
 \item TRSs $R$ and $R'$ are \emph{$\Fd$-globally isomorphic}, written $R\cong_{\GFE}R'$,
 if there exists a global isomorphism $\phi$ that is $\Vd$-invariant.
 Map $\phi$ is then called a \emph{$\Fd$-global isomorphism}.
 \item TRSs $R$ and $R'$ are \emph{locally isomorphic}, written $R\cong_{\LE}R'$, if 
they are in normal form and there are 
term isomorphisms $\phi_i:(\Fd\cup \Vd)\to (\Fd'\cup \Vd')$,
$1\le i\le n=|\Rd|$, such that 
$\phi(\Rd) = \{\phi_1(\ell_1)\to \phi_1(r_1),\ldots,\phi_n(\ell_n)\to \phi_n(r_n)\}
= \Rd'$,
where 
$\phi=(\phi_1,\ldots,\phi_n)$ denotes the family of such term isomorphisms.
Family $\phi$ is then called a \emph{local isomorphism}.
\item TRSs $R$ and $R'$ are \emph{$\Vd$-standard isomorphic}, written $R\cong_{\VSE}R'$,
if they are in $\Fd$-normal form, there is a family $\phi$ that is a local isomorphism and 
 $\phi_1|_{\Vd}=\ldots=\phi_n|_{\Vd}$. Family $\phi$ is then called a \emph{$\Vd$-standard isomorphism}.
\item TRSs $R$ and $R'$ are \emph{$\Fd$-standard isomorphic}, written $R\cong_{\SE}R'$, 
if they are in $\Vd$-normal form, there is a family $\phi$ that is a local isomorphism and 
 $\phi_1|_{\Fd}=\ldots=\phi_n|_{\Fd}$. Family $\phi$ is then called a \emph{$\Fd$-standard isomorphism}.
\item TRSs $R$ and $R'$ are \emph{$\Vd$-locally isomorphic}, written $R\cong_{\LVE}R'$, if they are  $\Fd$-standard isomorphic, and there is a $\Vd$-standard isomorphism such that the $\phi_i$'s are  $\Fd$-invariant. We then call $\phi$ a $\Vd$-local isomorphism.

 \item TRSs $R$ and $R'$ are  \emph{$\Fd$-locally isomorphic}, written $R\cong_{\LFE}R'$,  if they are  $\Vd$-standard isomorphic, and there is a $\Fd$-standard isomorphism such that the $\phi_i$'s are  $\Vd$-invariant. We then call $\phi$ a $\Fd$-local isomorphism.
\end{enumerate}
\end{defi}
Note that the definition of TRS isomorphisms with local properties in \cref{defi:TRSiso} is based on a fixed (arbitrary) ordering of rewriting rules $\Rd$.
A summary concerning all types of TRS isomorphism can be found in \cref{tab:TRSisos}. 
We differentiate between local, standard and global TRS isomorphisms, depending on the degree of local renamings involved.
By extending global TRS isomorphisms to constant families, $\cong_{\GE}$, $\cong_{\GFE}$ and $\cong_{\GVE}$ can also be understood as special cases of $\cong_{\LE}$.
This is in particular of good use when composing different types of TRS isomorphisms.

  \begin{ex}
    Consider the following TRSs:
    $$\begin{array}{l@{~}c@{~}l@{~}l@{~}l}
      \newR{trs-2}&=&(\{x,y\},&\{f,g,c\},&\{f(x,y) \to  c, g(x) \to  x\})\\
      \refR{trs-2}'&=& (\{x,y,z\},&\{f,g,c\},&\{f(x,y) \to  c, g(x) \to  x\})\\
      \newR{trs-3} &=& (\{z_1,z_2\},&\{f,g,c\},&\{f(z_1,z_2) \to  c, g(z_1) \to  z_1\})\\
      \newR{trs-4} &=&(\{x,y\},&\{f,g,c\},&\{g(x,y) \to  c, f(x) \to  x\})\\
      \newR{trs-5} &=& (\{z_1,z_2\},&\{f,g,c\},&\{g(z_1,z_2) \to  c, f(z_1) \to  z_1\})\\
      \newR{trs-6}&=&(\{z_1,z_2,z_3\},&\{f{,}g{,}c\},&\{g(z_1,z_2) \to  c{,} f(z_3) \to  z_3\})\\
      \newR{trs-7}&=&(\{x,y\}, &\{f,g,c,d\}, &\{f(x,y)  \to  c, g(x)  \to  c\})\\
      \newR{trs-8}&=&(\{z_1,z_2\}, &\{h,k,c,d\}, &\{h(z_1,z_2)  \to  c, k(z_1)  \to  d\})\\
      \newR{trs-9} &=&(\{x,y\},&\{f,g,h\},&\{f(x,y) \to g(y,x), f(x,y) \to h(x,y)\})\\
      \newR{trs-10}&=&(\{z_1,z_2\},&\{f,g,h\},&\{h(z_1,z_2) \to g(z_2,z_1), g(z_2,z_1) \to f(z_2,z_1)\})\\
      \newR{trs-11} &=& (\{z_1,z_2\},&\{f,g,h\},&\{f(z_1,z_2) \to g(z_2,z_1), f(z_2,z_1) \to h(z_1,z_2)\})\\
      \newR{trs-12} &=& (\{x,y\},&\{f,g,h\},&\{h(x,y) \to g(y,x), g(x,y) \to f(x,y)\}).
      \end{array}
      $$
    We have $\refR{trs-2} \cong_\GVE \refR{trs-3}$ (but not $\refR{trs-2} \cong_\GFE \refR{trs-3}$), since we have to rename
    variable $x$ into $z_1$ and variable $y$ into $z_2$ in both rules, which yields an 
    $\Vd$-global isomorphism.
    We have $\refR{trs-2} \cong_\GFE \refR{trs-4}$ by renaming $f$ into $g$ and vice versa in both rules, which consequently yields an $\Fd$-global isomorphism.
    Similar for
    $\refR{trs-5}$, 
    where $\refR{trs-2} \cong_\GE \refR{trs-5}$ (but $\refR{trs-2} \not\cong_\GVE \refR{trs-5}$ and $\refR{trs-2} \not\cong_\GFE \refR{trs-5})$.
    
    For $\refR{trs-6}$,
    neither \mbox{$\refR{trs-2}\cong_\GE \refR{trs-6}$,} $\refR{trs-2}\cong_\GFE \refR{trs-6}$ nor $\refR{trs-2}\cong_\GVE \refR{trs-6}$ holds,
    since incompatible cardinalities of variable sets
    $\{x,y\}$ and $\{z_1,z_2,z_3\}$ prevent the definition of appropriate term isomorphisms.
    Not even extending $\refR{trs-2}$'s variable set to $\{x,y,z\}$, \ie, working with 
    $\refR{trs-2}'$ changes this situation.
    Variable $x$ in $\refR{trs-2}'$ would still have to correspond to both $z_1$ \emph{and} $z_3$ in the rules of $\refR{trs-6}$. This is where the local property of standard TRS isomorphism helps: we have $\refR{trs-2}' \cong_\SE \refR{trs-6}$,
    since renaming of function symbols is global (mapping $f$ to $g$ and vice versa), but renaming of variables differs in each rule (mapping $x$ to $z_1$, $y$ to $z_2$ for the first rules and
    mapping $x$ to $z_3$ for the third rule).
    The idea is the same for $\Vd$-standard isomorphisms, they are global on the variables but allow for local renamings of function symbols.
    E.g., $\refR{trs-7}\cong_\VSE \refR{trs-8}$
    by renaming variables $x,y$ globally into $z_1,z_2$, 
    renaming  $f$ to $h$ in the first and $c$ to $d$ in the second rule.
    
    The local TRS isomorphisms are not restricted to global renaming of one or more symbol sets, but are satisfied with renaming of variables and/or function symbols on a per rule basis.
    E.g.~$\refR{trs-9}\cong_\LE \refR{trs-10}$, $\refR{trs-9} \not\cong_\LVE \refR{trs-10}$, 
    $\refR{trs-9}\not\cong_\LFE \refR{trs-10}$, $\refR{trs-7} \cong_\LVE \refR{trs-11}$ and $\refR{trs-7} \cong_\LFE \refR{trs-12}$.
    \end{ex}

\begin{remark}
Similar to our approach to term isomorphisms, one can first define TRS homomorphisms as a family of term homomorphisms of the same signature.
These term homomorphisms can then be imposed with additional restrictions to achieve an alternative characterisation of  \cref{defi:TRSiso}.
\end{remark}
Normal forms are helpful and were introduced
to prevent ``shrinking'' of rewriting rule sets and to ensure symmetry and transitivity of the isomorphism-relation in the first place.
Consider rewriting rule sets 
$$
\begin{array}{l@{~}c@{~}l}
\newRd{trs-13}&=&\{f(x)\to c,~g(x)\to d,~g(c)\to d\}\\
\newRd{trs-14}&=&\{f(x)\to d,~g(c)\to d\}\\
\end{array}
$$
where the  rewriting rules $f(x)\to c$, $g(x)\to d$ and $f(x)\to d$ are $\Fd$-equivalent. The induced TRSs $R_i = (\Fd_i,\Vd_i,\Rd_i)$ are $\Fd$-locally isomorphic in just one direction, \ie~one finds $\Fd$-invariant term isomorphisms $\phi=(\phi_1,\phi_2,\phi_3)$ such that 
$\phi(\refRd{trs-13})= \refRd{trs-14}$, but not the other way around due to incompatible cardinalities of said rule sets.
We will loosen up this restriction later in \cref{subsec:compatibility} when discussing templates, which present a natural connection between local TRS isomorphisms and normal forms.

The following lemma justifies the notion of ``isomorphisms'' between TRSs.

\begin{lemma}
\label{lem:equrel}
Every binary relation $\sim$, defined by $R \sim R'$ iff $R$ and $R'$ are \mbox{($\Vd$-/$\Fd$-)}locally/globally/ standard isomorphic TRSs, is an equivalence relation.
\end{lemma}
\begin{proof}
Fix TRSs $R_A,R_B,R_C$ such that $R_A\cong_{\LE} R_B$ and $R_B\cong_{\LE} R_C$ via local isomorphisms $\phi$ and $\phi'$ resp., and denote with 
$\Rd_x=\{\ell_1^{x}\to r_1^{x},\ldots,\ell_{n_x}^{x}\to r_{n_i}^{x}\}$,
the respective rule sets, where $n_x=|\Rd_x|$ for $x=A,B,C$.
Recall \cref{coro:termisos}.
We claim that $R_A$, $R_B$ and $R_C$ have the same number of rewriting rules, $n_A=n_B=n_C=n$.
Indeed, if $\ell_i^{A}\to r_i^{A},\ell_j^{A}\to r_j^{A}$, $1\le i<j\le n_A$ with 
$\phi_i(\ell_i^{A})\to \phi_i(r_i^{A})=\phi_j(\ell_j^{A})\to \phi_j(r_j^{A}),$
then for term isomorphisms $\psi_i:=\phi_j^{-1}\circ \phi_i$, $\psi_i(\ell_i^{A})\to \psi_i(r_i^{A})=\ell_j^{A}\to r_j^{A}$, 
in contradiction to $R_A$ in normal form.
The same holds for $R_B$.
{W.l.o.g.}~assume $R_A,R_B,R_C$ to be ordered such that 
$\phi_i(\ell_i^{A})\to \phi_i(r_i^{A})=\ell_i^{B}\to r_i^{B}$ 
and
$\phi_i'(\ell_i^{B})\to \phi_i'(r_i^{B})=\ell_i^{C}\to r_i^{C}$
for all $1\le i\le n$.
Then $R_A\cong_{\LVE}R_A$ and $R_B\cong_{\LVE}R_A$ by choosing $\id_{\Fd_A\cup \Vd_A}$ and $\phi^{-1}=(\phi_1^{-1},\ldots,\phi_n^{-1})$.
Moreover, $R_A\cong_{\LVE}R_C$ due to $\psi(\Rd_A)=\Rd_C$, where $\psi=(\phi_1'\circ \phi_1,\ldots,\phi_n'\circ \phi_n)$ is the
 (generalised)
  composition of $\phi$ and $\phi'$.
The remaining cases follow analogously by imposing additional restrictions to the term isomorphisms, which are then retained under composition and inversion.
\end{proof}

\begin{prop}
\label{prop:implications}
By \cref{defi:TRSiso}, $\Fd$-invariance implies $\Fd$-globality implies $\Fd$-locality and $\Vd$-invariance implies $\Vd$-globality implies $\Vd$-locality, and thus the following (strict) implications hold:
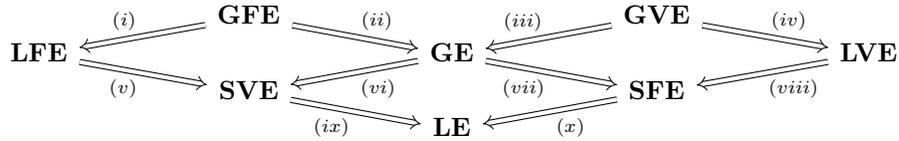
\begin{equation*}
\begin{tikzcd}[row sep=0pt, column sep=48pt]
& \GFE\arrow[Rightarrow]{dl}[swap]{(i)}\arrow[Rightarrow]{dr}{(ii)} && \GVE\arrow[Rightarrow]{dl}[swap]{(iii)}\arrow[Rightarrow]{dr}{(iv)}\\
\LFE \arrow[Rightarrow]{dr}[swap]{(v)} & & \GE \arrow[Rightarrow]{dl}{(vi)}\arrow[Rightarrow]{dr}[swap]{(vii)} & & \LVE \arrow[Rightarrow]{dl}{(viii)}\\
& \VSE \arrow[Rightarrow]{dr}[swap]{(ix)} & & \SE\arrow[Rightarrow]{dl}{(x)}\\
& & \LE
\end{tikzcd}
\end{equation*}
\end{prop}
\begin{proof}
The implications 
can be directly 
read from \cref{tab:TRSisos}.
Strictness 
follows from the examples provided in \cref{table-examples}, where
only rewriting rule sets $\Rd,\Rd'$ are given, and 
variables, function symbols and arities are omitted.
\begin{table}
\begin{tabular}
  {|@{}r@{\,}|@{\,}l@{\,}|@{\,}l@{\,}|@{\,}l@{\,}|@{\,}l@{\,}|}
\hline
&$\Rd$ & $\Rd'$ & Then\ldots & But\ldots
\\
\hline
$(i)$ &$\{f(x,y){\to} c, g(x){\to} c\}$ 
      &$\{f(x,y){\to} c, g(x){\to} d\}$
      &$R\cong_{\LFE} R'$
      &$R\not\cong_{\GFE} R'$
      \\
$(ii)$&$\{f(x,y){\to} c\}$
      &$\{f(y,x){\to} c\}$
      &$R\cong_{\GE}R'$
      &$R\not\cong_{\GFE}R'$
\\
$(iii)$&$\{f(x,y){\to} c, g(x){\to} c\}$ 
       &$\{f(x,y){\to} c, h(x){\to} c\}$
       &$R\cong_{\GE}R'$
       &$R\not\cong_{\GVE}R'$
\\
$(iv)$ &$\{f(x,y){\to} c, g(x){\to} c\}$
       &$\{f(y,x){\to} c,g(x){\to} c\}$
       &$R\cong_{\LVE}R'$
       &$R\not\cong_{\GVE}R'$
\\
$(v)$  &$\{f(x,y){\to} c, g(x){\to} c\}$ 
       &$\{f(y,x){\to} c, g(y){\to} d\}$
       &$R\cong_{\VSE}R'$
       &$R\not\cong_{\LFE}R'$
\\
$(vi)$ &$\{f(x,y){\to} c, g(x){\to} c\}$
       &$\{f(y,x){\to} c, g(y){\to} d\}$
       &$R\cong_{\VSE}R'$
       &$R\not\cong_{\GE}R'$
\\
$(vii)$ &$\{f(x,y){\to} c, g(x){\to} c\}$
        &$\{f(x,y){\to} c, g(y){\to} c\}$
        &$R\cong_{\SE}R'$
        &$R\not \cong_{\GE}R'$
\\
$(viii)$ &$\{f(x){\to} c\}$
         &$\{g(x)\to c\}$
         &$R\cong_{\SE}R'$
         &$R\not\cong_{\LVE}R'$
\\
$(ix)$   &$\{f(x,y){\to} c, g(x){\to} c\}$
         &$\{f(y,x){\to} c, g(x){\to} c\}$
         &$R\cong_{\LE}R'$
         &$R\not\cong_{\VSE}R'$
\\
$(x)$    &$\{f(x,y){\to} c, g(x){\to} c\}$
         &$\{f(y,x){\to} c, g(x){\to} d\}$
         &$R\cong_{\LE}R'$
         &$R\not\cong_{\SE}R'$
\\
\hline
\end{tabular}     
\caption{Examples showing strictness of implications in \cref{prop:implications}\label{table-examples}}
\end{table}
\end{proof}



\subsection{Semantic Properties of TRS Isomorphisms\label{subsec:semantic}}

We analyse w.r.t.~our hierarchy of TRS isomorphisms whether syntactically equivalent TRSs are
also semantically equivalent. Hence, in this section, we briefly analyse whether rewriting, convertibility, and termination of TRSs is preserved by the TRS isomorphisms.

  \begin{defi}[Term Isomorphism-Semantics]
    Let $\phi:(\Fd\cup \Vd)\to (\Fd'\cup \Vd')$ be a term isomorphism.
We apply $\phi$ also to contexts $C$ 
    in $T(\Fd,\Vd)$
    where $\phi([\cdot]) = [\cdot])$.
    For a substitution $\sigma=\{x_1 \mapsto t_1,\ldots,x_n \mapsto t_n\}$ with 
    domain $\Vd$ and codomain $T(\Fd,\Vd)$,
    denote with $\phi(\sigma)= \{\phi(x_1)\mapsto \phi(t_1),\ldots,\phi(x_n)\mapsto \phi(t_n)\}$ a new substitution w.r.t. $\phi$.
    These notions can analogously be extended to global TRS isomorphisms.
  \end{defi}
  Term isomorphisms and substitutions are then compatible in the sense that $\phi(C[\sigma(t)]) = \phi(C)[\phi(\sigma)(\phi(t)))]$ for any term $t\in T(\Fd,\Vd)$ and context $C$.
 
  Compatibility subsequently extends to global TRS isomorphisms: 


\begin{lemma}\label{prop:semantic-global}
Let $R=(\Fd,\Vd,\Rd),R'=(\Fd',\Vd',\Rd')$ be TRSs and $\phi$ be a global isomorphism from $R$ to $R'$ witnessing $R\cong_\GE R'$, $R \cong_\GFE R'$, or $R\cong_\GVE R'$. Then for any terms $s,t \in \T(\Fd,\Vd)$: $s \to_R t$ iff $\phi(s) \to_{\phi(R)} \phi(t)$. 
\end{lemma}
\begin{proof}
If 
$\ell\,{\to}\,r\in \Rd$, 
$s{=}C[\sigma(\ell)]\,{\to_R}\,C[\sigma(r)]{=}t$ with
$\sigma {=} \{x_1 {\mapsto} t_1,\ldots,x_n {\mapsto} t_n\}$, then 
 $\phi(s)=\phi(C[\sigma(\ell)]) = 
 \phi(C)[\sigma'(\phi(\ell))]
 \to_{\phi(R)} \phi(C)[\sigma'(\phi(r))]
 = \phi(C[\sigma(r)]) =  \phi(t)$ where $\sigma'=\phi(\sigma)$.

If $\phi(s) \to_{\phi(R)} \phi(t)$, then there exist $\ell'\,{\to}\,r' {\in} \phi(R)$, $\sigma' = \{x_1' {\mapsto} t_1', \ldots, x_n' {\mapsto} t_n'\}$ and a context $C'$ such that
$\phi(s) = C'[\sigma'(\ell')]$ and $\phi(t) = C'[\sigma'(r')]$.
Since $\phi$ is a global isomorphism, there exists $\phi^{-1}$ such that
$s = \phi^{-1}(C'[\sigma'(\ell')])$, $t = \phi^{-1}(C'[\sigma'(r')])$, 
$\phi^{-1}(C'[\sigma'(\ell')])=\phi^{-1}(C')[\sigma''(\phi^{-1}(\ell'))])$,
and $\phi^{-1}(C'[\sigma'(r')]) = \phi^{-1}(C')[\sigma''(\phi^{-1}(r'))])$  
where $\sigma''=\phi^{-1}(\sigma')$.
Clearly, $\phi^{-1}(\ell') \to \phi^{-1}(r') \in R$ since $\phi$ is a global isomorphism and thus $s \to_R t$.
\end{proof}

For TRS isomorphisms that are non-constant families of term isomorphisms (i.e.~local and standard ones), it is more complicated since one can apply a specific isomorphism of the family or any isomorphism of the family. However,
if only variables are locally renamed, this is still compatible with rewriting:

\begin{lemma}
 Let $R=(\Fd,\Vd,\Rd)$, $R'=(\Fd',\Vd',\Rd')$ be TRSs, $\phi=(\phi_1,\ldots,\phi_n)$ be a $\Vd$-local  (but not $\Fd$-local) TRS isomorphism  witnessing \mbox{$R \cong_\SE R'$,} or \mbox{$R\cong_\LVE R'$,} resp. Then for all $s,t \in \T(\Fd,\Vd)$ and all $1 \leq i \leq n$:
 $s \to_R t$ iff $\phi_i(s) \to_{\phi(R)} \phi_i(t)$. 
\end{lemma}
\begin{proof}
For the first direction, let $\ell \to r\in \Rd$, 
$s=C[\sigma(\ell)] \to_R C[\sigma(r)]=t$ with
$\sigma = \{x_1 \mapsto t_1,\ldots,x_n \mapsto t_n\}$ and $\sigma'=\phi|_\Fd(\sigma)$ where $\phi|_\Fd = \phi_i|_\Fd$ for $1 \leq i \leq n$.
Assume that $\phi$ maps $\ell \to r$ to $\phi_j(\ell)  \to \phi_j(r)$.
Then 
 $\phi_i(s)=\phi_i(C[\sigma(\ell)]) = \phi_i(C)[\sigma'(\phi_i(\ell))]$ 
and 
 $\phi_i(t)=\phi_i(C[\sigma(r)]) = \phi_i(C)[\sigma'(\phi_i(r)))]$.
 Since all $\phi_1,\ldots,\phi_n$ are global on $\Fd$, we can write the terms as follows:
 \begin{equation*}
  \begin{array}{l@{~}c@{~}l}
    \phi_i(s) & = & \phi_i(C)[(\sigma' \circ {\phi_i}|_\Vd \circ {\phi_j}^{-1}|_\Vd)(\phi_j(\ell)))],
    \\
    \phi_i(t) & = & \phi_i(C)[(\sigma' \circ {\phi_i}|_\Vd \circ {\phi_j}^{-1}|_\Vd)(\phi_j(r)))].
  \end{array}
 \end{equation*}
 Since $(\sigma' \circ {\phi_i}|_\Vd \circ {\phi_j}^{-1})$ is a substitution from variables to terms, we have $\phi_i(s) \to_{\phi(R)} \phi_i(r)$.

For the other direction, let $\phi_i(s) \to_{\phi(R)} \phi_i(t)$. Then there exist $\ell_j' \to r_j' \in \phi(R)$, with $\ell_j' = \phi_j(\ell_j), r_j'=\phi_j(r_j), \ell_j \to r_j \in R$,
$\sigma' = \{x_1' \mapsto t_1', \ldots, x_m' \mapsto t_m'\}$ and a context $C'$ such that
$\phi_i(s) = C'[\sigma'(\ell_j')]$ and $\phi_i(t) = C'[\sigma'(r_j')]$.
Note that $\phi_k$ can be written as $\phi_k|{_V} \opl \phi|_\Fd$, with constant $\phi|_\Fd:\Fd\to \Fd'$ independent of $1\le k\le n$, and that since all $\phi_k$ are bijective, the inverse $\phi_k^{-1}$ exists and can be written as $\phi_k^{-1}=\phi_k^{-1}|_\Vd \opl \phi^{-1}|_\Fd$.
We have $s = \phi_i^{-1}(C'[\sigma'(\ell_j')])=\phi_i^{-1}(C')[\sigma''({{\phi_j}|_\Vd}(\ell_j))])$
and $t = \phi_i^{-1}(C'[\sigma'(r_j')])= \phi_i^{-1}(C')[\sigma''({{\phi_j}|_\Vd}(r_j))])$
where 
\begin{equation*}
  \sigma'' = 
 \{
 {\phi_i^{-1}}|_\Vd(x_1') \mapsto 
 {\phi_i^{-1}}(t_1'),\ldots, 
 {\phi_i^{-1}}|_\Vd(x_n') \mapsto 
 {\phi_i^{-1}}(t_n')
\}.
\end{equation*}
Since $\sigma'' \circ {\phi_j|_\Vd}$ is a substitution from variables to terms and $\ell_j \to r_j \in R$, we have
$s \to_R t$.
\end{proof}
The two previous lemmas imply that global and $\Vd$-local isomorphisms (that are $\Fd$-global) are compatible with rewriting, convertibility, and termination:

\begin{theo}[Semantic Compatibility of Global/$\Vd$-Local TRS Isomorphisms]\label{theo:semantic:compatibility}
 Let $R=(\Fd,\Vd,\Rd),R'=(\Fd',\Vd',\Rd')$ 
 such that 
 $R\cong_\GE R'$, $R \cong_\GFE R'$, $R\cong_\GVE R'$, $R \cong_\SE R'$, or $R \cong_\LVE R'$.
 Let $\phi$ be the isomorphism witnessing the corresponding equivalence.
 
 Then the following holds:
 \begin{itemize}
 \item[$(i)$] 
 $s \to_R t$ iff $\phi(s)\to_{R'} \phi(t)$ 
 where in case of a $\Vd$-local equivalence, $\phi$ is any isomorphism $\phi_i$ of the family $(\phi_1,\ldots,\phi_n)$.
\item[$(ii)$] 
 $s \xleftrightarrow{*}_R t$ iff $\phi(s)\xleftrightarrow{*}_{R'} \phi(t)$ where in case of a $\Vd$-local equivalence, $\phi$ is any isomorphism $\phi_i$ of the family $(\phi_1,\ldots,\phi_n)$.
  \item[$(iii)$] $R$ is terminating iff $\phi(R)$ is terminating.
\end{itemize}
\end{theo}

For $\Fd$-local TRS isomorphism, the situation changes. Only a weak correspondence, as formulated in the following proposition, holds:

\begin{prop}[Semantic Properties of $\Fd$-Local TRS Isomorphism]\label{prop:semantic:f-local}
Let $R=(\Vd,\Fd,\Rd),R'=(\Vd',\Fd',\Rd')$ be TRSs such that 
$R \cong_\VSE R'$,  $R \cong_\LFE R'$, or $R \cong_\LE R'$. 
Let $\phi=(\phi_1,\ldots,\phi_n)$ be the ($\Fd$-local) isomorphism witnessing the equivalence of $R$ and $R'$
such that
$\Rd = \{\ell_1 \to r_1,\ldots,\ell_n \to r_n\}$,
$\Rd' = \phi(R)= \{\phi_1(\ell_1) \to \phi_1(r_1) ,\ldots,\phi_n(\ell_n) \to \phi_n(r_n)\}$.
Then the following holds:
\begin{enumerate}
\item[$(i)$] If $s \to_R t$, then there exists $1 \leq i \leq n$ such that $\phi_i(s) \to_{\phi(R)} \phi_i(t)$.
However, $R,R',\phi$ can be chosen such that there exist $s,t,i$ with  $s \to_R t$, but  $\phi_i(s) \not\to_{\phi(R)} \phi_i(t)$.

\item[$(ii)$] If $\phi_i(s) \to_{\{\phi_i(\ell_i) \to \phi_i(r_i)\}} \phi_i(t)$, then $s \to_R t$.
However, $R,R',\phi,j$ can be chosen such that there exist with  $\phi_j(s) \to_{R'}
\phi_j(t)$, but $s \not\to_R t$.

\item[$(iii)$]  $R,R'$ can be chosen such that $R$ is terminating, but $R'$ is not terminating.
\end{enumerate}
\end{prop}
\begin{proof}
  $(i)$: For the first part, 
    choose $\phi_i$ such that $\ell_i \to r_i$ is the rewrite rule that is applied in the step $s \to_R t$. Then $\phi_i$ behaves like the global isomorphism in the proof of \cref{prop:semantic-global} which shows the claim. 
    For the second part, we provide a counter-example: 
    Let 
    $\Fd = \Fd' = \{f,g,h,u\}$, $\Vd = \Vd'$, 
    $$\begin{array}{l@{~}c@{~}l}
       \Rd&=&\{f(x) \to g(x,x), f(x) \to h(x,x,x)\},\\
       \Rd'&=& \{f(x) \to g(x,x), u(x) \to h(x,x,x)\}
      \end{array}
    $$
and $\phi=(\phi_1,\phi_2)$
with ${\phi_1}|_\Fd = \id_\Fd$ and ${\phi_2}|_\Fd = \{f \mapsto u, h \mapsto h, g\mapsto g, u \mapsto h\}$.
Then $f(x) \to_R h(x,x,x)$, but $\phi_1(f(x)) = f(x) \not\to_{R'} h(x,x,x) = \phi_1(h(x,x,x))$.

%
$(ii)$: The first part holds, since the inverse of $\phi_i$ can be applied to the terms $\phi_i(s)$ and $\phi_i(t)$, showing that $s \to_{\{\ell_i \to r_i\}} t$ and thus $s \to_R t$. 
For the second part, consider $\Fd=\Fd'=\{a,f,g\}$, $\Vd = \Vd'$,
    $$\begin{array}{l@{~}c@{~}l}
\Rd&=&\{f(x) \to g(x), g(x) \to g(a)\},\\
\Rd'&=&\{g(x) \to f(x), g(x) \to g(a)\}
\end{array}
$$
and
${\phi_1}|_\Fd =\{f \mapsto g, g\mapsto f, a\mapsto a\}$,
${\phi_2}|_\Fd = \id_\Fd$,
${\phi_i}|_\Vd = \id_\Vd$ for $i=1,2$.
Then $\phi_1(f(x))= g(x) \to_{R'} g(a) = \phi_1(f(a))$,
but $f(x) \not\to_R f(a)$.

$(iii)$: Consider $\Fd=\Fd'=\{a,f,g,h\}$, $\Vd = \Vd'$,
    $$\begin{array}{l@{~}c@{~}l}
\Rd&=&\{f(x) \to g(x), h(x) \to g(a)\},\\ 
\Rd'&=&\{f(x) \to g(x), g(x) \to f(a)\}
\end{array}
$$
and 
${\phi_1}|_\Fd =\id_\Fd$,
${\phi_2}|_\Fd = \{h \mapsto g, g\mapsto f, f\mapsto h\}$,
${\phi_i}|_\Vd = \id_\Vd$ for i=1,2.
Then $R$ is terminating, since it replaces all $f$ and $h$ by $g$, and $g$ cannot be rewritten,
but $R'$ is nonterminating due to $f(a) \to_{R'} g(a) \to_{R'} f(a)$.\qedhere
\end{proof}

\subsection{Discussion and Applications}
\label{subsect:application}
Consider
$\newR{trs-15}=(\refFd{trs-15},\{x,y\},\refRd{trs-15}))$ where $\refFd{trs-15}=\{c,s,f\}$ and  
$$\refRd{trs-15}=\{f(c,x)\to x, f(s(x),y)\to s(f(x,y))\}.$$
This system can be interpreted as a single recursive definition of addition on natural numbers by equations
$\add(0,m) =  m$ and 
$\add(\suc(n),m)  = \suc(\add(n,m))$:
TRS $\refR{trs-15}$ is $\Fd$-standard isomorphic to the TRS 
$\newR{trs-16}=(\refFd{trs-16},\refVd{trs-16},\refRd{trs-16})$ where 
$\refFd{trs-16}=\{0,\suc,\add\}$, $\refVd{trs-16}=\{n,m\}$, and 
$$\refRd{trs-16}=\{\add(0,m)\to m, \add(\suc(n),m)\to \suc(\add(n,m))\}$$ 
courtesy of 
$$\begin{array}{l@{~}c@{~}l}
\phi_1=\{c\mapsto 0, s\mapsto \suc, f\mapsto \add, x\mapsto m, y\mapsto n\}~\text{and}\\
\phi_2=\{c\mapsto 0, s\mapsto \suc, f\mapsto \add, x\mapsto n, y\mapsto m\}.
 \end{array}
$$
Our interpretation only holds if we (explicitly via term isomorphisms or not) {\rename} $f$ globally, \ie~the $f$ in the second rule has to refer to the same function as in the first.
In other words, functions are seen as global quantities, while variables are treated as local placeholders.
In this sense, $\Fd$-standard isomorphism is rightly named as the in practice most important type of TRS isomorphism.  

In summary, TRS isomorphisms that permit to rename variables locally and function symbols globally seem to be quite natural following the principle that  ``names of formal parameters do not matter'', \ie, whether a TRS has a rule  $f(x) \to x$  or $f(y) \to y$ should not make a difference, independently whether variables $x$ and $y$ occur in other rules.

However, often the signature is fixed, and one does not want to rename the function symbols. For instance,  in Knuth-Bendix-completion, one compares single rules without renaming of function symbols, \ie~$\cong_{\LVE}$ (on singleton TRSs) is the relevant equivalence notion.

If local renamings are involved, we require the TRS to be in normal form. 
This is not necessary if only global renamings are used, which for instance allows identifying
 TRSs as equal even if they have ``renamed'' rules, thus all global TRS isomorphism may have practical applications in this case.

However, for the TRS isomorphisms that allow local renaming of function symbols, we cannot provide a use case where they can be used as a proper notion of equality. The results in \cref{subsec:semantic} also support this judgement. However, besides including all dimensions in the hierarchy of TRS isomorphisms, there are two justifications to analyze the notions, both require that the notions are in $\Pc$ (which holds as we show in \cref{subsec:localTRSinP}): 
(1) They provide a fast (but incomplete) test to refute the corresponding standard or global equivalences, using the contrapositive of the implications in \cref{prop:implications}. 
(2) In addition, they can also be used to construct a polynomial (but incomplete) test to refute graph isomorphism since we provide polynomial translations from graphs to TRSs in \cref{globalisos}.
Efficiently implemented algorithms could thus aid in (sophisticated) common subexpression elimination by refuting if selected code-blocks are isomorphic, \ie~if they offer the same functionality after a suitable renaming of functions or methods and variables.
Since deciding global and standard equivalences is $\GI$-complete, it is interesting to consider related problems that are in $\Pc$. This may help to understand the inclusions
$\Pc \subseteq \GI \subseteq \NP$.

\section{\texorpdfstring{Templates and Local Properties}{Templates and Local Properties}\label{sect:templates}}

We state two explicit polynomial-time algorithms which solve $\LVE$, $\LFE$, $\LE$, and later serve to justify polynomial reductions from $\SE$ to $\GFE$ and $\VSE$ to $\GVE$.
This method is known as the \emph{template-method} (an example concerning structural isomorphisms on context-free grammars can be found in ~\cite{rosenkrantz-hunt:85}) and involves \emph{canonical renaming} of variables or function symbols or both on a per rule basis, depending on the type of local TRS isomorphism.
For a fixed TRS $R=(\Fd,\Vd,\Rd)$, 
let $\sz_\Rd=\max\{|\ell|,|r|: \ell\to r\in \Rd\}$ be the maximum length of terms appearing in $\Rd$.
For the sake of simplicity, we assume terms or rewriting rules are stored in a computationally advantageous structure, \eg, as trees, where each symbol can be accessed in at least linear time, i.e. $\Od(\sz_\Rd)$.
Furthermore, symbols themselves can be represented as hash maps which allows access in $\Od(\log|\Vd|)$ or $\Od(\log|\Fd|)$ respectively.

\subsection{Templates and Maximal Normal Forms}
\begin{defi}[Standardised Symbol Sets and Templates]
\label{defi:templates}
Let $R=(\Fd,\Vd,\Rd)$ be a TRS, $\Rd=\{\ell_1\to r_1,\ldots,\ell_n\to r_n\}$, and fix $1\le j\le n$.
Set $\Vd^S=\{x_1,\ldots,x_K\}$, $K=|\Vd|$, and $\Fd^S=\{f_{k,l}: l\in L, 1\le k\le p_l\}$ where $L=\{\ar(f): f\in \Fd\}$, $p_l=|\{f\in \Fd: \ar(f)=l\}|$ and $|\Fd^S|=|\Fd|$.
We call $\Vd^S$ the \emph{standardised variable set} and $\Fd^S$ the \emph{standardised function symbol set} of $R$.
We define a local renaming $\nu_j: (\Fd\cup \Vd)\to (\Fd\cup \Vd^S)$ of only variables the following way: 
Consider $\ell_j$ and replace each occurrence of a variable with $x_k$, where $x_k$ refers to the $k$th distinct variable in $\ell_j$.
Since $\Var(r_j)\subset \Var(\ell_j)$ this is already sufficient to state an $\Fd$-invariant term isomorphism.
Analogously, \rename~function symbols locally via $\mu_j: (\Fd\cup \Vd)\to (\Fd^S\cup \Vd)$ by replacing each occurrence of a function symbol with $f_{k,l}$, where $f_{k,l}$ is the $k$th distinct function symbol in $\ell_j \to r_j$ of arity $l$ ($k$ depends on $l$), resulting in a $\Vd$-invariant term isomorphism.
We call $\nu_i(\ell_i)\to \nu_i(r_i)$ the \emph{$\Vd$-template} and  $\mu_i(\ell_i)\to \mu_i(r_i)$ the \emph{$\Fd$-template} of rewriting rule $\ell_i\to r_i$.
Moreover, $\nu=(\nu_1,\ldots,\nu_n)$ and $\mu=(\mu_1,\ldots,\mu_n)$ are called the \emph{$\Vd$-/$\Fd$-template isomorphism} of $R$, respectively.
Let  $\pi=(\pi_1,\ldots,\pi_n)$ where
$\pi_i=(\mu_i|_\Fd\opl \id_{\Vd^S}) \circ \nu_i$ (which is well-defined due to $\Fd$-invariance of $\mu_i$ for each $1\le i\le n$). We  call $\pi_i(\ell_i)\to \pi_i(r_i)$ 
the \emph{template} of rewriting rule $\ell_i\to r_i\in \Rd$ and $\pi$ the \emph{template isomorphism} of $R$.
The notion of the ($\Vd$-/$\Fd$-)template is then extended 
to $\Rd$ ($\nu(\Rd),\mu(\Rd),\pi(\Rd)$), called \emph{\mbox{($\Vd$-/$\Fd$-)}template set}, and
to all of TRS $R$ ($\nu(R),\mu(R),\pi(R)$), yielding new TRSs by combining \mbox{($\Vd$-/$\Fd$-)}template sets and (possibly) standardised symbol sets.
\end{defi}
Note that in the above definition strictly speaking $\nu$ and $\mu$ (and in turn $\pi$) not only depend on $\Vd$ or $\Fd$ but also on $\Rd$ (and $\ar$), i.e. $\nu=\nu^R,\mu=\mu^R,\pi=\pi^R$.
Only $\Vd^S$ solely depends on the initial set $\Vd$, $\Fd^S$'s uniqueness additionally requires $\ar$ and in turn $R$, i.e. $\Fd^S=\Fd^{S,R}$. 
These dependencies are are to be inferred from the respective context and are explicitly stated if ambiguous.

\begin{lemma}\label{lemma:template-computation}
($\Vd$-/$\Fd$)-templates of TRSs can be computed in polynomial time.
\end{lemma}
\begin{proof}
\cref{algo:Vtemplate,algo:Ftemplate} compute $\nu$, $\mu$, and standardised symbol sets in a combined at worst runtime of $\Od(\sz_\Rd|\Rd||\Vd||\Fd|\log(|\Vd||\Fd|))$ for a TRS $R=(\Fd,\Vd,\Rd)$ by linearly parsing over each rewriting rule.
$\Vd$-/$\Fd$-template sets are then the result of polynomial map-application of $\nu$ or $\mu$ respectively.
Combination of both yields the general case.
\end{proof}
As an example, recall \cref{ex:TRSHom}.
Then the standardised sets are $\Vd^S=\{x_1,x_2\}$ and $\Fd^S=\{f_{1,0},f_{1,1},f_{2,1},f_{1,3}\}$.
The $\Vd$-template of $f(g(x),y)\to h(x)$ is $f(g(x_1),x_2)\to h(x_1)$
and the $\Fd$-template is $f_{1,2}(f_{1,1}(x),y)\to f_{2,1}(x)$.
Extend this to rule set $\{f(g(x),y)\to h(x), g(f(x,y))\to c\}$.
Now we see why double indices in the functions symbol set are needed to ensure consistent arities. 
Renaming function symbols in the same pattern as variables would yield $\Fd$-template set
 $\{f_1(f_2(x),y)\to f_3(x), f_1(f_2(x,y))\to f_3\}$ which can not be constructed out of one singular term set due to both $f_1$ and $f_2$ appearing with arity $1$ and $2$.
Moreover, it is not sufficient to just trivially increase the first index, \ie~remapping $f(g(x),y)\to h(x)$ to $f_{1,2}(f_{2,1}(x),y)\to f_{3,1}(x)$.
The resulting $\Fd$-template set would be 
$\{f_{1,2}(f_{2,1}(x),y)\to f_{3,1}(x), f_{1,1}(f_{2,2}(x,y))\to f_{3,0}\}$,
 which requires an underlying function symbol set of at least order six, namely $\{f_{1,1},f_{1,2},f_{2,1},f_{2,2},f_{3,0},f_{3,1}\}$.
This prevents bijectivity of $\mu_j$'s, and thus them being well-defined term isomorphisms.

\begin{lemma}
\label{lem:maximalform}
Every TRS $R=(\Fd,\Vd,\Rd)$ can be brought into a \emph{maximal \mbox{($\Vd$-/$\Fd$-)} normal form} $R'=(\Fd,\Vd,\Rd')$, where $\Rd'\subset \Rd$ and for every $\Rd'\subsetneq \Rd''\subset \Rd$, $(\Fd,\Vd,\Rd'')$ is not in \mbox{($\Vd$-/$\Fd$-)}normal form.
This \mbox{($\Vd$-/$\Fd$-)}normal form is unique up to \mbox{($\Vd$-/$\Fd$-)}equivalence of rewriting rules (and in turn symbol sets) and constructible in polynomial time.
\end{lemma}
\begin{proof}
Let $R=(\Fd,\Vd,\Rd)$ be a TRS, $\Rd=\{\ell_1\to r_1,\ldots,\ell_n\to r_n\}$, and fix $1\le j\le n$.
We consider $\Vd$-normal forms and $\Vd$-templates.
We first compute the  {$\Vd$-template} of each rewriting rule $\ell_i\to r_i$.
$\Vd$-equivalence of distinct rewriting rules $\ell\to r,\ell'\to r'$ is equivalent to $\Vd$-local isomorphism of corresponding singleton rule sets $\{\ell\to r\}$ and $\{\ell'\to r'\}$.
Corresponding equivalence classes can 
be determined once we have the $\Vd$-template isomorphism $\nu$, since by \cref{lem:equrel}, 
$[\ell_i\to r_i]=\{\ell_j\to r_j\in \Rd:\nu_i(\ell_i)\to \nu_i(r_i)=\nu_j(\ell_j)\to \nu_j(r_j)\}.$
Thus, $\nu(\Rd)$ directly implies a partitioning of $\Rd$ into $1\le m\le n$ sets of rewriting rules of the same $\Vd$-template.
Then every representation system $\Rd'=\{\ell_{i_1}\to r_{i_1},\ldots,\ell_{i_m}\to r_{i_m}\}$ of this equivalence relation is already in maximal $\Vd$-normal form.
Note that choosing minimal symbol sets is also permissible but would lead to incompatible cardinalities.
This is why the symbol sets are for now stated to remain unchanged (something that will be adressed in \cref{subsec:compatibility} and \cref{defi:genTRSiso}).

We omit the reasoning for $\Fd$-normal forms, since it is completely analogous.
For normal forms, 
we sequentially apply \cref{algo:Vtemplate,algo:Ftemplate}, \ie~we take the maximal $\Fd$-normal form of the maximal $\Vd$-normal form of $R$. 
\end{proof}

\begin{algorithm}[t]
\DontPrintSemicolon
\SetKwFunction{Sort}{Sort}
\SetKwFunction{TempName}{TempName}
\SetKwFunction{RemoveDuplicates}{RemoveDuplicates}
\SetKwFunction{Insert}{Insert}
\SetKwFunction{TempFunc}{TempFunc}
\SetKwFunction{map}{map}

\Input{TRS $R=(\Fd,\Vd,\ar,\Rd)$ where $\Rd=\{\ell_1\to r_1,\ldots,\ell_n\to r_n\}$}
\Output{$\nu_{\Vd}=(\nu_1,\ldots,\nu_n)$ family of $\Fd$-invariant term isomorphisms, canonical renaming of variables on a per rule basis}
\Runtime{$\Od(\sz_\Rd|\Rd||\Vd||\Fd|\log(|\Vd||\Fd|))$}
Initialise new variable set $\Vd^S\leftarrow\{x_1,\ldots,x_{|\Vd|}\}$\;
\For(\tcp*[f]{$\Od(\sz_\Rd|\Rd|\log|\Vd|)$}){$j\leftarrow 1$ \KwTo $n$} {
  Initialise partial map $\nu_j: \Fd\cup \Vd\to \Fd\cup \Vd^S$\;
  Let 
  $\gamma_1\ldots\gamma_m = \ell_j\in (\Fd\cup \Vd)^*$
  \tcp*[l]{store terms without symbols $($, $)$, $,$}
  Set $k\leftarrow 1$\;
  
  \For(\tcp*[f]{rename variables from left to right $\Od(\sz_\Rd\log|\Vd|)$}){$i\leftarrow 1$ \KwTo $m$} {
    \uIf(\tcp*[f]{$\Od(\log|\Vd|)$}){$\gamma_i\in \Vd$ and $\nu_j(\gamma_i)$ undefined} {
      Set $\nu_j(\gamma_i)\leftarrow x_k$\;
      $k\leftarrow k+1$\;
    }
  }
}
\tcp{extend $\nu_j$ to all of $\Fd\cup\Vd$}
  \For(\tcp*[f]{$\Od(|\Vd|\log|\Vd|)$}){$x\in \Vd$} {
    \uIf(\tcp*[f]{$\Od(\log|\Vd|)$}){$\nu_j(x)$ undefined} { 
      Set $\nu_j(x)\leftarrow x_k$\;
      $k\leftarrow k+1$\;
    }
  }
  \lFor(\tcp*[f]{$\Od(|\Fd|\log|\Fd|)$}){$f\in \Fd$} {
    Set $\nu_j(f)\leftarrow f$
  }
 \Return $\nu=(\nu_1,\ldots,\nu_n)$\;
 \caption{Computation of $\Vd$-template 
 }
 \label{algo:Vtemplate}
\end{algorithm}

\begin{algorithm}[t]
\DontPrintSemicolon
\SetKwFunction{Sort}{Sort}
\SetKwFunction{TempName}{TempName}
\SetKwFunction{RemoveDuplicates}{RemoveDuplicates}
\SetKwFunction{Insert}{Insert}
\SetKwFunction{TempFunc}{TempFunc}
\SetKwFunction{map}{map}

\Input{TRS $R=(\Fd,\Vd,\ar,\Rd)$ where $\Rd=\{\ell_1\to r_1,\ldots,\ell_n\to r_n\}$}
\Output{$\mu=(\mu_1,\ldots,\mu_n)$ family of $\Vd$-invariant term isomorphisms, canonical renaming of function symbols on a per rule basis}
\Runtime{$\Od(\sz_\Rd|\Rd||\Vd||\Fd|\log(|\Vd||\Fd|))$}
\lFor(\tcp*[f]{$\Od(|\Fd|\log |\Fd|)$}){$l\in \{\ar(f):f\in \Fd\}$}{
  let $p_l\leftarrow 0$ }
Initialise new function symbol set $\Fd^S\leftarrow\emptyset$\;
\For(\tcp*[f]{$\Od(\sz_\Rd|\Rd|\log|\Fd|)$}){$j\leftarrow 1$ \KwTo $n$} {
  Initialise partial map $\mu_j: \Fd\cup \Vd\to \Fd^S\cup \Vd$\;
  Let
  $\gamma_1\ldots\gamma_m = \ell_jr_j\in (\Fd\cup \Vd)^*$
  \tcp*[l]{store terms without symbols $($, $)$, $,$}
  \For(\tcp*[f]{rename function symbols}\tcp*[f]{$\Od(\sz_\Rd\log|\Fd|)$}){$i\leftarrow 1$ \KwTo $m$} {
    \uIf{$\gamma_i\in \Fd$ and $\mu_j(\gamma_i)$ undefined} {
      Let $l\leftarrow \ar(\gamma_i)$ \tcp*[r]{$\Od(\log |\Fd|)$}
      Set $p_l\leftarrow p_l+1$\;
      Set $\Fd^S\leftarrow\Fd^S\cup \{f_{p_l,l}\}$\;
      Set $\mu_j(\gamma_i)\leftarrow f_{p_l,l}$ \tcp*[r]{$\Od(\log |\Fd|)$}
    }
  }
}
\tcp{extend $\mu_j$ to all of $\Fd\cup\Vd$}
  \For(\tcp*[f]{$\Od(|\Fd|\log|\Fd|)$}){$f\in \Fd$} {
    \uIf{$\mu_j(f)$ undefined} { 
      Let $l\leftarrow \ar(f)$ \tcp*[r]{$\Od(\log |\Fd|)$}
      Set $p_l\leftarrow p_l+1$\;
      Set $\Fd^S\leftarrow\Fd^S\cup \{f_{p_l,l}\}$\;
      Set $\mu_j(\gamma_i)\leftarrow f_{p_l,l}$ \tcp*[r]{$\Od(\log |\Fd|)$}
    }
  }
  \lFor(\tcp*[f]{$\Od(|\Vd|\log|\Vd|)$}){$x\in \Vd$} {
    Set $\mu_j'(x)\leftarrow x$
  }
 \Return $\mu=(\mu_1,\ldots,\mu_n)$\;
 \caption{Computation of $\Fd$-template 
 }
 \label{algo:Ftemplate}
\end{algorithm}

\subsection{\texorpdfstring{Local TRS Isomorphisms are~in $\Pc$}{Local TRS Isomorphisms are in P}}
\label{subsec:localTRSinP}
We will show that local TRS isomorphisms 
$\LE$, $\LVE$ and $\LFE$ are in $\Pc$. 
A necessary first step is the following corollary: 

\begin{coro}
\label{coro:templateiso}
A TRS $R$ in \mbox{($\Vd$-/$\Fd$-)}normal form is \mbox{($\Vd$-/$\Fd$-)}locally isomorphic to its \mbox{($\Vd$-/$\Fd$-)}template.
Moreover, as direct consequence of \cref{lem:equrel}, two TRSs $R_i=(\Fd_i,\Vd_i,\Rd_i)$, $i=1,2$, in \mbox{($\Vd$-/$\Fd$-)}normal form are \mbox{($\Vd$-/$\Fd$-)}locally isomorphic iff they have the same \mbox{($\Vd$-/$\Fd$-)}template
set and the same standardised symbol sets, \ie~$\Vd_1^S=\Vd_2^S$ and $\Fd_1^S=\Fd_2^S$.
\end{coro}

\begin{theo}[Local TRS Isomorphisms are~in $\Pc$]
\label{theo:localisoinp}
Isomorphisms $\LE$, $\LVE$ and $\LFE$ can be decided in polynomial time.
\end{theo}
\begin{proof}
By \cref{coro:templateiso}, two TRSs $R_i=(\Fd_i,\Vd_i,\Rd_i)$ in \mbox{($\Vd$-/$\Fd$-)}normal form are \mbox{($\Vd$-/$\Fd$-)}locally isomorphic iff their \mbox{($\Vd$-/$\Fd$-)}template sets $\Rd_1'$ and $\Rd_2'$ are the same and their corresponding standardised symbol sets agree.
Both construction of \mbox{($\Vd$-/$\Fd$-)}templates and verification of set equality can be done in polynomial time, with latter being possible in at most most $\Od(\sz_{\Rd_1}|\Rd_1|\log |\Rd_1|+\sz_{\Rd_2}|\Rd_2|\log |\Rd_2|)$.
\end{proof}

%
%

\subsection{Delocalisation of Standard TRS Isomorphisms}
\label{subsec:delocalisation}
In this section, we explain how local TRS isomorphisms can be used to transform standard TRS isomorphisms into syntactically equivalent global TRS isomorphisms
by \emph{delocalisation}.
 Only $\Fd$-standard isomorphisms remain semantically equivalent to their $\Fd$-global counterpart (\cref{theo:semantic:compatibility} and \cref{prop:semantic:f-local}).

\begin{lemma}[Delocalisation of Standard TRS Isomorphisms]
\label{lem:localisation}
$\Vd$-/$\Fd$-standard isomorphism of TRSs is equivalent to $\Vd$-/$\Fd$-global isomorphism of their respective $\Fd$-/$\Vd$-templates.
\end{lemma}
\begin{proof}
Fix TRSs $R_i=(\Fd_i,\Vd_i,\Rd_i)$, $i=1,2$, in $\Vd$-normal form and consider $\Vd$-templates $R_i'=(\Fd_i,\Vd_i^S,\nu^ {R_i}(\Rd_i))$ 
as generated 
in polynomial time by 
\cref{algo:Vtemplate}.
We claim that $R_1\cong_{\SE}R_2$ iff $R_1'\cong_{\GFE}R_2'$.
In both cases, \cref{coro:templateiso} and existence of valid term isomorphisms guarantee that all four rule sets $\Rd_i$, $\nu^{R_i}(\Rd_i)$ posses the same number of rules $n$, and that $\Vd_1^S=\Vd_2^S=\Vd^S$ (either by $\Vd$-invariance of $\Fd$-global isomorphism, or due to $|\Vd_1|=|\Vd_2|$).

Let $\phi=(\phi_1,\ldots,\phi_n)$ be a $\Fd$-standard isomorphism witnessing $R_1\cong_{\SE}R_2$.
Then $\phi'=(\phi'_1,\ldots,\phi'_n)$, $\phi'_i = (\nu^{R_2}_i)\circ \phi_i\circ({\nu^{R_1}_i})^{-1}$, yields an $\Fd$-global isomorphism between $R_1'$ and $R_2'$.
Indeed, for fix $1\le i\le n$, $\phi'_i$ is bijective,
$
\phi'_i(\Vd^S)=(\nu^{R_2}_i\circ \phi_i\circ(\nu^{R_1}_i)^{-1})(\Vd^S)
=(\nu^{R_2}_i\circ \phi_i)(\Vd_1)
= \nu^{R_2}_i(\Vd_2)
= \Vd^S
$
and $\phi'_i|_{\Fd_1}=\phi_i|_{\Fd_1}$, due to $(\nu^{R_1}_i)^{-1}|_{\Fd_1}=\id_{\Fd_1}$, $\nu^{R_2}_i|_{\Fd_2}=\id_{\Fd_2}$, as long as we assume  $\Rd_i$, $\nu^{R_i}(\Rd_i)$
 to be ordered in such a way that we can viably compose those functions.
Moreover, every term isomorphism $\phi_i'$ is $\Vd$-invariant due to construction in \cref{algo:Vtemplate} and $\Vd$-equivalence of the images of the $i$th rewriting rule $\ell_i\to r_i\in \Rd_1$ under $\phi_i|_{\Fd_1}\opl \id_{\Vd_1}$ and $\phi_i$, resulting in $\phi'_1=\ldots=\phi'_n$, \ie~$\phi'$ is an $\Fd$-global isomorphism.
Then $\phi'(\nu^{R_1}(\Rd_1))=(\nu^{R_2}\circ \phi\circ(\nu^{R_1})^{-1})(\nu^{R_1}(\Rd_1))=\nu^{R_2}(\Rd_2)$ and we are done.
For given $\Fd$-global isomorphism $\phi'=(\phi'_1,\ldots,\phi'_n)$ between $R_1'$ and $R_2'$ one shows the same way that 
$\phi:= ((\nu^{R_2}_1)^{-1}\circ\phi'_1\circ\nu^{R_1}_1,\ldots,(\nu^{R_2}_n)^{-1}\circ\phi'_n\circ\nu^{R_1}_n)$ defines an $\Fd$-standard isomorphism between $R_1$ and $R_2$.
That is to say, Diagram (\ref{eq:FSEtoGFEDia}) is commutative. 
Indeed, $\phi_i|_{\Fd_1}=\phi'_i|_{\Fd_1}$ due to $\Fd$-invariance of $(\nu^{R_1})^{-1}$ and $\nu^{R_2}$,
 and thus $\phi_1|_{\Fd_1}=\ldots=\phi_n|_{\Fd_1}$.
\begin{equation}
\begin{tikzcd}
(\Fd_1,\Vd_1,\Rd_1)\arrow{rrrrrr}{\phi\text{ $\Fd$-standard isomorphism}}\arrow[shift left=0.75ex]{d}{\nu^{R_1}}& & & & & & (\Fd_2,\Vd_2,\Rd_2)\arrow[shift left=0.75ex]{d}{\nu^{R_2}}
\\
\big(\Fd_1,\Vd^S,\nu^{R_1}(\Rd_1)\big)\arrow[shift left=0.75ex]{rrrrrr}{\phi'\text{ $\Fd$-global isomorphism}}\arrow[shift left=0.75ex]{u}{(\nu^{R_1})^{-1}} & & & & & & \big(\Fd_2,\Vd^S,\nu^{R_2}(\Rd_2)\big)\arrow[shift left=0.75ex]{u}{(\nu^{R_2})^{-1}}
\end{tikzcd}
\label{eq:FSEtoGFEDia}
\end{equation}
Analogously, TRSs $R_i=(\Fd_i,\Vd_i,\Rd_i)$ in $\Fd$-normal form, $i=1,2$, are $\Vd$-standard isomorphic iff their $\Fd$-templates $R_i'=(\Fd_i,\Vd_i,\mu^{R_i}(\Rd_i))$ as given by \cref{defi:templates} and \cref{algo:Ftemplate}, are $\Vd$-globally isomorphic.
We only have to argue that both $\Fd$-templates posses the same standardised function symbol set $\Fd^S$ if initial TRSs $R_i$ are $\Vd$-standard isomorphic, but this immediately clear due to existence of a term isomorphism between $T(\Fd_1,\Vd_1)$ and $T(\Fd_2,\Vd_2)$.
Indeed, in this case $\{\ar_1(f): f\in \Fd_1\}=\{\ar_2(f): f\in \Fd_2\}$ and 
$|\ar_1^{-1}(l)|=|\ar_2^{-1}(l)|$ for any $l\in \N_0$ (refer to \cref{defi:templates}).
\end{proof}

  \begin{remark}
    \label{rem:rsubseteql}
    Recall the notations from \cref{sect:trs}.
    Without the assumptions $\Var(r)\subseteq \Var(\ell)$ and/or $\ell$ not a variable for any rewriting rule $\ell\to r$ in some TRS $R$, 
    \cref{algo:Vtemplate} and \ref{algo:Ftemplate} are still polynomial.
    \cref{algo:Vtemplate} would have to be altered to also scan the right side $r$ for appearances of variables not yet in $\ell$ with an additional runtime of only $\Od(\sz_\Rd\log|\Vd|)$.
     \cref{algo:Ftemplate} is unaffected.
    In particular, all results of \cref{sect:templates} remain true for TRSs without these assumptions.
\end{remark}

\subsection{Generalisation of Standard and Local TRS Isomorphisms}
\label{subsec:compatibility}
The proposed TRS isomorphisms can be seen as
core equalities where more involved equalities can be derived from. We discuss such a generalisation in this section.
Recall that finiteness and minimality of symbol sets were 
essential
 assumptions when first introducing TRSs.
They aspire to closely mirror practical application and eliminate noise in the form of unnecessary symbols.
This however leads to incompatibility of certain symbol sets when comparing selected TRSs.
Consider rewriting rules 
$$\begin{array}{l@{~}c@{~}lll@{~}c@{~}l}
  \newRd{trs-18}&=&\{f(x)\to c, f(x)\to h(x)\} &\text{and} &\newRd{trs-19}&=&\{f(x)\to c, g(x)\to h(x)\}\end{array}$$ implicitly extended to TRSs $R_i=(\Fd_i,\Vd_i,\Rd_i)$.
Clearly, both posses the same $\Fd$-template set 
$\{f_{1,1}(x)\to f_{1,0},f_{1,1}(x)\to f_{2,1}(x)\}$ and one would consider them to be $\Fd$-locally isomorphic.
Despite this heuristic intuition, we cannot find valid term isomorphisms on underlying term sets $T(\refFd{trs-18},\refVd{trs-18})$ and $T(\refFd{trs-19},\refVd{trs-19})$, since $\refFd{trs-18}=\{f,h\}$ and $\refFd{trs-19}=\{f,g,h\}$ do not have the same number of function symbols, which is essential for bijectivity of underlying term homomorphisms.
This is no problem however.
Note, that $\refFd{trs-18}^S=\{f_{1,0},f_{1,1},f_{2,1}\}\subseteq \{f_{1,0},f_{1,1},f_{2,1},f_{3,1}\}=\refFd{trs-19}^S$.
Now artificially extend $\refFd{trs-18}$ to $\refFd{trs-18}'=\refFd{trs-18}\cup \refFd{trs-19}^S\setminus \refFd{trs-18}^S$ by one unary ``dummy'' function symbol $f_{3,1}$.
Then $\refR{trs-18}'=(\refFd{trs-18}',\refVd{trs-18},\refRd{trs-18})$ is a valid TRS-extension of $\refR{trs-18}$ and $\refR{trs-18}'\cong_{\LFE}\refR{trs-19}$.
This procedure can  be generalised to handle arbitrary TRSs as long as they posses the same $\Fd$-template sets, simply by comparing the standardised function symbol sets (which 
can be done in polynomial time).
The same concept also holds for ($\Vd$-)templates.
In fact, if any TRSs $R_i=(\Fd_i,\Vd_i,\Rd_i)$, $i=1,2$, posses the same \mbox{($\Vd$-/$\Fd$-)}template set, then already 
$\Fd_1^S\subseteq \Fd_2^S$ (or $\Fd_2^S\subseteq \Fd_1^S$) independent of the type of template, while
set-relation $\Vd_1^S\subseteq \Vd_2^S$ (or $\Vd_2^S\subseteq \Vd_1^S$) holds trivially by design.
Therefore finding appropriate term isomorphisms to extend local TRS isomorphisms based on template-defining ones (refer to \cref{defi:templates} and Algorithms \ref{algo:Vtemplate} and \ref{algo:Ftemplate}), is reduced to mathematical technicalities, involving appropriate set- and map-extensions in addition to concatenation.
The same holds true for standard TRS isomorphisms, where an extension on the locally renamed symbol set leaves the global-property of the initial standard TRS isomorphism on the complementary symbol set unaffected.

Furthermore, the ($\Fd$-/$\Vd$-)normal form-requirement
in the definition of standard/local TRS isomorphisms can be relaxed by comparing maximal \mbox{($\Vd$-/$\Fd$-)} normal forms, since
by \cref{lem:maximalform}, computing ($\Fd$-/$\Vd$-)normal forms is intertwined with computing ($\Fd$-/$\Vd$-)templates.
Together with
\cref{coro:templateiso} and \cref{lem:localisation} this gives rise to the 
the following extended definition of TRS isomorphisms, which is based on comparing ($\Fd$-/$\Vd$-)templates:
\begin{defi}[Generalisation of TRS Isomorphisms]
\label{defi:genTRSiso}
We say that two TRSs $R_1,R_2$ are generalised-\mbox{($\Vd$-/$\Fd$-)}locally isomorphic if the \mbox{($\Vd$-/$\Fd$-)}template sets of the maximal \mbox{($\Vd$-/$\Fd$-)}normal forms of $R_1$ and $R_2$ are the same.
Moreover, if the $\Vd$-/$\Fd$-templates 
(with minimal symbol sets)
 of the maximal $\Vd$-/$\Fd$-normal forms of $R_1$ and $R_2$ are $\Fd$-/$\Vd$-globally isomorphic, we say that $R_1,R_2$ are generalised-$\Fd$-/$\Vd$-standard isomorphic.
Denote isomorphisms with $\cong_{\LE^*}$, $\cong_{\LFE^*}$, $\cong_{\LVE^*}$, $\cong_{\SE^*}$, and $\cong_{\VSE^*}$, respectively.
\end{defi}

\section{\texorpdfstring{$\GI$-Completeness of Standard and Global TRS Isomorphisms}{GI-Completeness of Standard and Global TRS Isomorphisms}\label{globalisos}}

It turns out that global, non-invariant, renaming of at least one symbol set, immediately results in $\GI$-completeness of the corresponding TRS isomorphism-decision problem:

\begin{theo}[$\GI$-Completeness of TRS Isomorphisms] 
\label{theo:TRSisos}
The following sets are polynomially equivalent:
\begin{itemize}
 \item[$(i)$] $\GI=\{(G,G'): G,G'\text{ isomorphic graphs}\}$
 \item[$(ii)$] $\GVE=\{(R,R'): R,R' \text{ $\Vd$-globally isomorphic TRS}\}$
 \item[$(iii)$] $\GFE=\{(R,R'): R,R' \text{ $\Fd$-globally isomorphic TRS}\}$
 \item[$(iv)$] $\GE=\{(R,R'): R,R' \text{ globally isomorphic TRS}\}$
 \item[$(v)$] $\SE=\{(R,R'): R,R' \text{ $\Fd$-standard isomorphic TRS}\}$
 \item[$(vi)$] $\VSE=\{(R,R'): R,R' \text{ $\Vd$-standard isomorphic TRS}\}$
 \end{itemize}
\end{theo}

\subsection{Encoding of TRSs into Outgoing-Ordered Labelled Directed Graphs}
In preparation for proving \cref{theo:TRSisos}, we discuss an encoding from TRSs into so-called \emph{Outgoing-Ordered Labelled Directed Graphs}.
For the rest of the paper, 
assume $\Fd\cup \Vd$ and $\N_0\cup\{T,F,C,g,z\}$ to be disjoint (symbol) sets.

Our graph model of choice is the following:
A \emph{labelled directed graph (LDG)} is a tuple $G=(V,E,L,\lab)$, where $V$ is a finite set of vertices, $E \subseteq V\times V\times L$ are directed labelled 
edges between vertices, $L$ is a finite set of labels, 
and $\lab:V\to L$ is a labelling 
function.
Vertex $v\in V$ is called a \emph{root}, if it has no incoming edges.
Graph $G$ is called \emph{connected}, 
if we can find a path between any two vertices in the undirected graph corresponding to $G$.
Vertex $v\in V$ is called \emph{initial}, if every other vertex $w\in V\setminus \{v\}$ is reachable from $v$, 
\ie~there exists a path from $v$ to $w$.
We 
call $G$ \emph{rooted}, if $G$ is connected and has a unique, initial root. 
If additionally 
there is at most one path between any two vertices in the undirected graph corresponding to $G$, then
$G$ is a \emph{tree}.
Two LDGs $G_i=(V_i,E_i,L_i,\lab_i)$ are \emph{isomorphic} \wrt~one-to-one correspondence $\rho: L_1\to L_2$, $G_1\cong G_2$ ($\rho$ denoted indirectly), if we find a 
bijection $\phi:V_1\to V_2$ 
such that adjacency and label-equivalence classes are preserved, \ie~$(v,w,l)\in E_1$ iff 
$(\phi(v),\phi(w),\rho(l))\in E_2$, and $\rho(\lab_1(v))=\lab_2(\phi(v))$ for $v\in V_1$.
Map $\phi$ is 
called a \emph{graph isomorphism}.
If both label sets agree and $\rho$ can be chosen as identity between them, we call $G_1$ and $G_2$ \emph{strongly isomorphic}, $G_1\cong_S G_2$,
and $\phi$ a \emph{strong graph isomorphism}.
In this case, we omit $\rho$ entirely in our notation.

We can canonically encode a term into a tree by respecting the term structure.
E.g. $f(x,y)$ is represented by a tree
 with $f$-labelled root and two child-vertices, labelled with $x$ and $y$ respectively.
However, the order of arguments is of utmost importance, since function symbols are ranked (we do not want to treat terms $f(x,y)$ and $f(y,x)$ as isomorphic). That is why we use OOLDG-trees:
An \emph{outgoing-ordered labelled directed graph (OOLDG)} \cite{schmidt-schauss-et-al:2013} is a 
special case of an LDG,
where edge-labels of outgoing edges are unique, \ie~if $(w,v,l),(w,v',l)$ are included edges, then already $v=v'$.

For the proof of \cref{theo:TRSisos}, 
we will use the complexity classes derived from (strong) isomorphisms on directed graphs or OOLDGs, resp.
They are polynomial equivalent to $\GI$, which was for instance proved in \citep{booth-colbourn:77,schmidt-schauss-et-al:2013}.
\begin{prop}[$\GI$-Completeness of Certain Graph-Classes, see 
\citep{booth-colbourn:77} and \citep{schmidt-schauss-et-al:2013}]
\label{prop:OOLDGScomplete}
The following sets are polynomially equivalent:
\begin{itemize}
\itemsep0em
 \item[$(i)$] $\GI=\{(G,G'): G,G'\text{ isomorphic graphs}\}$
 \item[$(ii)$] $\DG=\{(G,G'): \text{$G,G'$ isomorphic unlabelled directed graphs}\}$
 \item[$(iii)$] $\OOLDG=\{(G,G'): \text{$G,G'$ strongly isomorphic OOLDGs}\}$
\end{itemize}
\end{prop}
In essence, we identify TRSs by a forest of unconnected OOLDG-trees with numbered edges, where every connected component directly corresponds to a single rewriting rule.
Depending on the type of the analysed TRS isomorphism, we  modify this forest by adding additional vertices and edges or replacing 
labels.
Of particular interest are $\GFE$, $\GVE$ and $\GE$, where we explicitly state such a polynomial reduction into OOLDGs.

We use the following construction: 
Suppose $T_1,\ldots,T_n$ are OOLDG-trees with distinct (labelled) roots $v_1,\ldots,v_n$.
Graph $\join(v,l,T_1,\ldots,T_n)$ is then the OOLDG-tree with $l$-labelled root $v$ and ordered subtrees $T_1,\ldots,T_n$, \ie~extend the union of $T_1,\ldots,T_n$ to a new OOLDG-tree by introducing fresh
edges $(v,v_i,i)$, $1\le i\le n$.
An example can be found in \cref{fig:termTreeForest}.
\begin{figure}[t]
\begin{minipage}[b]{.5\textwidth}
 \centering 
\begin{tikzpicture}[every node/.append style={thick,minimum size=0.7cm},yscale=.8,xscale=.6]
\node[shape=circle,draw=black] (v) at (0,0)  {$v$};

\node[shape=circle,draw=black] (v1) at (-4,-1.5)  {$v_1$};

\node[shape=circle,draw=black] (v2) at (4,-1.5)  {$v_n$};

\path [->,ultra thick,above] (v) edge node {$n$} (v2);
\path [->,ultra thick,above] (v) edge node {$1$} (v1);
\path [->,ultra thick,above,dotted] (v) edge node {} (-2,-1.5);
\path [->,ultra thick,above,dotted] (v) edge node {} (2,-1.5);

\node[] (T1) at (-4,-3.25)  {$T_1$};
\draw[dotted,ultra thick] (-4,-2) -- (-3,-4) -- (-5,-4) -- (-4,-2);

\node[] (T1) at (4,-3.25)  {$T_n$};
\draw[dotted,ultra thick] (4,-2) -- (5,-4) -- (3,-4) -- (4,-2);

\node[] (T1) at (0,-3.25)  {$\cdots$};

\end{tikzpicture}%
\caption{Visualisation of $\join(v{,}l{,}T_1{,}\ldots{,}T_n)$}
\label{fig:termTreeForest} 
\end{minipage}
\begin{minipage}[b]{.5\textwidth}
 \centering
\begin{tikzpicture}[every node/.append style={thick,minimum size=0.7cm},yscale=.8,xscale=.5]
\node[shape=rectangle,draw=black] (v) at (0,0)  {$T$};

\node[shape=circle,draw=black] (v111) at (-7,-4.5)  {$x$};
\node[shape=circle,draw=black,fill=gray!20] (v112) at (-5,-4.5)  {$y$};

\node[shape=circle,draw=black,fill=gray!40] (v11) at (-6,-3)  {$h$};
\node[shape=circle,draw=black] (v12) at (-2,-3)  {$x$};

\node[shape=circle,draw=black,fill=gray!60] (v1) at (-4,-1.5)  {$f$};

\node[shape=circle,draw=black] (v21) at (2,-3)  {$x$};
\node[shape=circle,draw=black,fill=gray!20] (v22) at (6,-3)  {$y$};

\node[shape=circle,draw=black,fill=gray!40] (v2) at (4,-1.5)  {$h$};

\path [->,ultra thick,above] 
 (v1) edge node[yshift=-2mm,xshift=-1mm] {$1$} (v11) 
 (v11) edge node[yshift=-2mm,xshift=-2mm] {$1$} (v111) 
 (v2) edge node[yshift=-2mm,xshift=-1mm] {$1$} (v21) 
;

\path [->,ultra thick,above]   
 (v2) edge node[yshift=-2mm,xshift=1mm] {$2$} (v22)
 (v1) edge node[yshift=-2mm,xshift=1mm] {$2$} (v12) 
 (v11) edge node[yshift=-2mm,xshift=2mm] {$2$} (v112)  
;

\path [->,ultra thick,above] (v) edge node[yshift=-1mm,xshift=1mm] {$2$} (v2);
\path [->,ultra thick,above] (v) edge node[yshift=-1mm,xshift=-1mm]{$1$} (v1);

\end{tikzpicture}%
\caption{Term tree $\Tree(f(h(x{,}y){,}x){\to}h(x{,}y))$}\label{fig:termTree}
\end{minipage}\end{figure}
For a term $t\in T(\Fd,\Vd)$, its \emph{term tree}  $\Tree(t)$
is an OOLDG-tree, inductively defined by:
If $t=x$ or $t=c$ for $x\in \Vd$ or $c\in \Fd$, then $\Tree(t)$ is a single vertex with respective label.
If $t=f(t_1,\ldots,t_{\ar(f)})$ for $f\in \Fd$ and terms $t_1,\ldots,t_{\ar(f)}\in T(\Fd,\Vd)$, then $\Tree(t)=\join(v,f,\Tree(t_1),\ldots,\Tree(t_{\ar(f)}))$, where $v$ is fresh.
For a rewriting rule $\ell\to r$, we generalise the definition to $\Tree(\ell\to r)=\join([\ell\to r],T,\Tree(\ell),\Tree(r))$, where $[\ell\to r]$ is a fresh $T$-labelled root (see \cref{fig:termTree} for an example).
For a TRS
$R=(\Fd,\Vd,\Rd)$, we call $G(R)=\bigcup_{\ell\to r\in \Rd}\Tree(\ell\to r)$, 
the \emph{TRS-forest} of $R$, where the union of graphs is understood as the union of their vertex/edge-sets and extension of their labelling functions.
It can be constructed in time 
$\Od(\sz_\Rd|\Rd|)$, where $s$ is an upper bound on the length of rewriting rules, if we, for example, use doubly linked lists by linearly parsing each rewriting rule $\ell\to r$.
The following lemmas highlight the correlation between (global) isomorphism of TRSs and isomorphism between corresponding TRS-forests.

\begin{lemma}
  \label{lem:TRStreeiso1to2}
  If TRSs $R_1$ and $R_2$ are \mbox{($\Vd$-/$\Fd$-)}globally isomorphic then TRS-forests $G(R_1)$ and $G(R_2)$ are isomorphic.
\end{lemma}
\begin{proof}
  Fix a global isomorphism $\phi:(\Fd_1 \cup \Vd_1)\to (\Fd_2 \cup \Vd_2)$ and
let $\Rd_1=\{\ell_1\to r_1,\ldots,\ell_n\to r_n\}$ and $\Rd_2=\{\ell_1'\to r_1',\ldots,\ell_n'\to r_n'\}$ be ordered with respect to $\phi$, \ie~$\phi_i(\ell_i)\to \phi(r_i)=\ell_i'\to r_i'$ for each $1\le i\le n$.
Then $\rho=\phi\opl \id_{\N\cup \{T\}}$ is well-defined and bijective.
The following OOLDG isomorphisms are understood to be w.r.t. $\rho$.
Let $T_i=\Tree(\ell_i\to r_i)$ and $T_i'=\Tree(\ell_i'\to r_i')$.
To define an OOLDG isomorphism $\phi':V_1\to V_2$, it suffices to consider restrictions $\phi_i':V(T_i)\to V(T_i')$ on connected components $T_i=\Tree(\ell_i\to r_i)$ and $T_i'=\Tree(\ell_i'\to r_i')$, where $V(G)$ for a  graph $G$ denotes its 
\emph{vertex set}.
Map $\phi'$ is given by $\bigsqcup_{i=1}^n \phi_i'$.
If $v_i$ and $v_i'$ are the roots of $T_i$ and $T_i'$ resp., set $\phi'_i(v_i)=v_i'$.
Now $\phi_i'$ can be broken down into maps $\phi_i^l:V(\Tree(\ell_i))\to V(\Tree(\ell_i'))$ and $\phi_i^r:V(\Tree(r_i))\to V(\Tree(r_i'))$, such that $\phi_i'=\{v\mapsto v'\}\opl \phi_i^l\opl \phi_i^r$.
It remains to show that if $\phi(t)=t'$ for terms $t\in T(\Fd_1,\Vd_1)$ and $t'\in T(\Fd_2,\Vd_2)$, indeed $\Tree(t)\cong \Tree(t')$.
We use structural induction.
Denote with $\phi_t$ an isomorphism between $\Tree(t)$ and $\Tree(\phi(t))$.

In the base case, either $t=x$ or $t=c$ for a variable $x\in \Vd_1$ or a constant $c\in \Fd_1$.
Thus $\phi(t)=\phi(x)$ or $\phi(t)=\phi(c)$, and $\Tree(t)$, $\Tree(\phi(t))$ consist of one vertex $v,v'$ resp., with corresponding label.
Then $\phi_t=\{v\mapsto v'\}$ is an isomorphism.
Now let $t=f(t_1,\ldots,t_n)$ for some  $f\in \Fd_1$ and terms $t_1,\ldots,t_n$.
Suppose $\Tree(t_i)\cong \Tree(\phi(t_i))$ via isomorphisms $\phi_{t_i}$, $1\le i\le n$, according to the induction hypothesis.
We have $\Tree(t)=\join(v,f,\Tree(t_1),\ldots,\Tree(t_n))$ and
$\Tree(\phi(t))=\join(v',\phi(f),\Tree(\phi(t_1)),\ldots,$ $\Tree(\phi(t_n)))$
due to $\phi(t)=\phi(f)(\phi(t_1),\ldots,\phi(t_n))$.
Then $\phi_t=\{v\mapsto v'\}\opl\bigsqcup_{i=1}^n \phi_{t_i}$ is a well-defined isomorphism between the term trees.
Bijection $\rho$ is additionally $\Fd$-/$\Vd$-invariant if 
TRS isomorphism $\phi$ is  $\Fd$-/$\Vd$-invariant (in case of $\Vd$-/$\Fd$-global isomorphism).
\begin{figure}[b]
\centering
\begin{tikzpicture}[every node/.append style={thick,minimum size=0.7cm},scale=0.8]

\node[shape=circle,draw=black] (v1) at (0,0)  {$v_i$};

\path [->,ultra thick,above] (v1) edge node {$2$} ( 2,-1.5);
\path [->,ultra thick,above] (v1) edge node {$1$} (-2,-1.5);

\node[] (T1) at (-2,-2.75)  {$\ell_i$};
\draw[dotted,ultra thick] (-2,-1.5) -- (-1,-3.5) -- (-3,-3.5) -- (-2,-1.5);

\node[] (T1) at (2,-2.75)  {$r_i$};
\draw[dotted,ultra thick] (2,-1.5) -- (3,-3.5) -- (1,-3.5) -- (2,-1.5);

\node[shape=circle,draw=black] (v2) at (7,0)  {$v_{i}'$};

\path [->,ultra thick,above] (v2) edge node {$1$} (9,-1.5);
\path [->,ultra thick,above] (v2) edge node {$2$} (5,-1.5);

\node[] (T1) at (5,-2.75)  {$r_{i}'$};
\draw[dotted,ultra thick] (5,-1.5) -- (6,-3.5) -- (4,-3.5) -- (5,-1.5);

\node[] (T1) at (9,-2.75)  {$\ell_{i}'$};
\draw[dotted,ultra thick] (9,-1.5) -- (10,-3.5) -- (8,-3.5) -- (9,-1.5);

\draw[thick] (-3.25,-1.25) -- (-0.75,-1.25) -- (-0.75,-3.75) -- (-3.25,-3.75) -- (-3.25,-1.25);
\draw[thick] (0.75,-1.25) -- (3.25,-1.25) -- (3.25,-3.75) -- (0.75,-3.75) -- (0.75,-1.25);
\draw[thick] (3.75,-1.25) -- (6.25,-1.25) -- (6.25,-3.75) -- (3.75,-3.75) -- (3.75,-1.25);
\draw[thick] (7.75,-1.25) -- (10.25,-1.25) -- (10.25,-3.75) -- (7.75,-3.75) -- (7.75,-1.25);

\draw [<->, bend angle=15, bend right,thick]  (2,-3.75) to node[below] {} node[below,sloped,xshift=5mm,yshift=1mm] {$\phi_i^r$} (5,-3.75);

\draw [<->, bend angle=17, bend right,thick]  (-2,-3.75) to node[below] {} node[below,sloped,yshift=6mm,xshift=35mm] {$\phi_i^l$} (9,-3.75);

\draw [<->, bend angle=10, bend left,thick]  (v1) to node[below] {} node[below,sloped,yshift=1mm] {$\phi_i'$} (v2);

\end{tikzpicture}%
\caption{Illustrating the reduction of isomorphisms on term trees 
in the proof of \cref{cor:TRStreeiso}}
\end{figure}
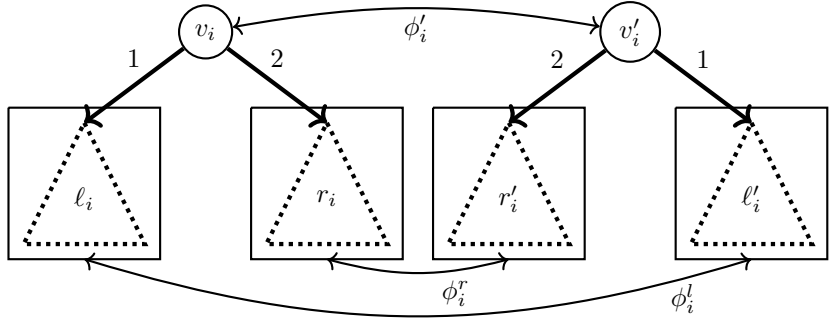   
\end{proof}

The reverse implication does not hold in general.
Both edge-label invariance and symbol-relationship preservation are necessary and non-trivial requirements to ensure TRS isomorphism.
E.g. consider singleton TRSs 
$$
\begin{array}{l@{~}c@{~}l}
\newR{trs-20}&=&(\{f,c\},\{x\},\{f(x,c)\to f(x,c)\})
\\
\newR{trs-21}&=&(\{f,c\},\{x\},\{f(c,x)\to f(c,x)\})
\end{array}
$$
Clearly, $\refR{trs-20}$ and $\refR{trs-21}$ cannot be isomorphic in any way due to different positionings of $x,c$ in both rewriting rules. 
In fact, any TRS isomorphism between $\refR{trs-20}$ and $\refR{trs-21}$ is equivalent to $f(x,c)\to f(x,c)$ and $f(c,x)\to f(c,x)$ being equivalent.
On the contrary, $G(\refR{trs-20})\cong G(\refR{trs-21})$ \wrt~one-to-one correspondences $\rho=\{x\mapsto c,c\mapsto x\}\opl\id_{\{1,2,f,T}\}$ or
$\rho'=\{1\mapsto 2,2\mapsto 1\}\opl\id_{\{x,c,f,T}\}$.
The former isomorphism is edge-invariant but not symbol-relationship preserving, the latter exactly the opposite.

\begin{lemma}
  \label{lem:TRStreeiso2to1}
  Suppose $G(R_1)$ and $G(R_2)$ are isomorphic for TRSs $R_1,R_2$.
  Then $R_1$ and $R_2$ are \mbox{($\Vd$-/$\Fd$-)}globally isomorphic if edge-labels are invariant and the symbol-relationship is preserved, \ie~if $G(R_1)$ and $G(R_2)$ are isomorphic \wrt~$\rho:L_1\to L_2$, then $\rho|_\N=\id_\N$ and $\rho(\Vd_1)=\Vd_2$ (and 
  $\rho(\Fd_1)=\Fd_2$) (for $\Vd$-/$\Fd$-global isomorphism, we demand $\rho$ to be invariant on $\Fd$/$\Vd$ only).
\end{lemma}
\begin{proof}
  Suppose 
$\phi':V_1\to V_2$ is an OOLDG isomorphism between $G(R_1)$ and $G(R_2)$ \wrt~$\rho:L_1\to L_2$, such that $\rho|_{\N}=\id_{\N}$ and $\rho(\Vd_1)=\Vd_2$.
Then $\rho(T)=T$ due to $\phi'$ mapping root vertices onto root vertices.
Moreover, $|\Rd_1|=|\lab_1^{-1}(\{T\})|=|\lab_2^{-1}(\{T\})|=|\Rd_2|$ and thus there is a one-to-one correspondence between term trees in $G(R_1)$ and $G(R_2)$.
Set $\phi=\rho|_{\Fd_1\cup \Vd_1}$, well-defined and bijective.
Note that $\ar_1(f)=\outdeg(v)$ for each vertex $v\in V_1$ if $\lab_1(v)=f\in \Fd_1$, and therefore, since $\phi'$ preserves outward degrees, $\ar_1(f)=\ar_2(\phi(f))$ for any function symbol $f\in \Fd_1$.
Map $\phi$ is a term isomorphism due to symbol-relationship preserving property of label set function $\rho$.

We now show that if $\phi'(\Tree(\ell\to r))=\Tree(\ell'\to r')$ for term trees $\Tree(\ell\to r)$, $\ell\to r\in \Rd_1$ and $\Tree(\ell'\to r')$, $\ell'\to r'\in \Rd_2$, then already $\phi(\ell)\to \phi(r)=\ell'\to r'$.
Isomorphism $\phi'$ necessarily maps $[\ell\to r]$ onto $[\ell'\to r']$ and thus, due to invariance of edge-labels, it suffices to check that $\phi(t)=t'$ if $\Tree(t)\cong \Tree(t')$ via $\phi'$, for any terms $t\in T(\Fd_1,\Vd_1)$ and $t'\in T(\Fd_2,\Vd_2)$.
This can be shown by structural induction over $t$.
In the base case, $t=x$ for some variable $x\in \Vd_1$, \ie~$\Tree(t)$ only consists of one vertex $v$.
Then $\Tree(t')$ is necessarily a single vertex $v'$ and due to $\rho(\Vd_1)=\Vd_2$, $\lab_2(v')\in \Vd_2$, \ie~$\phi(t)=\phi(x)=\rho(\lab_1(v))=\lab_2(v')=t'$.
Analogously if $t=c$ for some constant $c\in \Fd_1$.
Now suppose $t$ is of the form $t=f(t_1,\ldots,t_n)$ for some $n$-ary function symbol $f\in \Fd_1$ and terms $t_1,\ldots,t_n$, \ie~$\Tree(t)=\join(v,f,T_1,\ldots,T_n)$, where $T_i=\Tree(t_i)$, $1\le i\le n$.
Then $\Tree(t')=\join(v',f',T_1',\ldots,T_n')$, $t'=f'(t_1',\ldots,t_n')$, where $f'\in \Fd_2$ is an $n$-ary function symbol and $t_1',\ldots,t_n'$ are terms such that $T_i'=\Tree(t_i')$, $1\le i\le n$.
Consequently, $f'=\rho(f)=\phi(f)$ and $T_i\cong T_i'$ for any $i\le i\le n$ due to assumption on $\phi$ and $\rho$.
By induction hypothesis, $\phi(t_i)=t_i'$ and thus $\phi(t)=\phi(f(t_1,\ldots,t_n))=\phi(f)(t_1',\ldots,t_n')=f'(t_1',\ldots,t_n')=t'$.
Again, $\Fd$-/$\Vd$-invariance of $\rho$ results in 
$\Fd$-/$\Vd$-invariance of $\phi$, or equivalently, resulting TRS isomorphism $\phi$ is $\Vd$-/$\Fd$-global.
\end{proof}

The gradually constructed isomorphism $\phi'$ in the proof of \cref{lem:TRStreeiso1to2} is edge-label-invariant and preserves symbol-relationship, \ie~$R_1,R_2$ ($\Vd$-/$\Fd$-)globally isomorphic iff $G(R_1),G(R_2)$ are isomorphic under tightened label-assumptions.
As mentioned, the bijection $\rho=\phi\opl\id_{\N\cup\{T\}}$ is additionally $\Fd$-/$\Vd$-invariant if given global TRS isomorphism $\phi$ is  $\Fd$-/$\Vd$-invariant (in case of $\Vd$-/$\Fd$-global isomorphism).
We can thus combine Lemmas \ref{lem:TRStreeiso1to2} and \ref{lem:TRStreeiso2to1}:
\begin{coro}
\label{cor:TRStreeiso}
For TRSs $R_1,R_2$, consider the following  statements:
\begin{itemize}
  \item[$(i)$] $R_1$ and $R_2$ are \mbox{($\Vd$-/$\Fd$-)}globally isomorphic.
 \item[$(ii)$] $G(R_1)$ and $G(R_2)$ are isomorphic.
\end{itemize}
Then
$(i)\Longrightarrow(ii)$ holds.
The reverse implication is only true if edge-labels are invariant and the symbol-relationship is preserved, \ie~if $G(R_1)$ and $G(R_2)$ are isomorphic \wrt~$\rho:L_1\to L_2$, then $\rho|_\N=\id_\N$ and $\rho(\Vd_1)=\Vd_2$ (and 
$\rho(\Fd_1)=\Fd_2$) (for $\Vd$-/$\Fd$-global isomorphism, we demand $\rho$ to be invariant on $\Fd$/$\Vd$).
\end{coro}

\subsection{\texorpdfstring{Proof of \cref{theo:TRSisos}}{Proof of Theorem 5.1}}
We proceed with the proof of \cref{theo:TRSisos} by a chain of polynomial reductions depicted in \cref{fig:TRSisos} (where an arrow $A \Rightarrow B$ represents a reduction $A \preceq^\Pc B$).
\begin{figure}[th]    
\centering
\begin{tikzcd}[column sep=14pt,row sep=30pt]
%
(vi)\;\VSE                 
 &&(ii)\;\GVE  \arrow[Leftarrow]{ll}[sloped,above]{\text{Prop.\,\ref*{prop:FSEtoGFEinitiallylemma}}}
  & (iv)\;\GE \arrow[Rightarrow,bend left=15]{dd}[sloped,above]{\text{Prop.\,\ref*{prop:GEtoGI}}}\arrow[Leftarrow,bend right=15]{dd}[sloped,below]{\text{Prop.\,\ref*{prop:GItoTRS}}}
   &  (iii)\;\GFE  
    && (v)\;\SE \arrow[Rightarrow]{ll}[sloped,above]{\text{Prop.\,\ref*{prop:FSEtoGFEinitiallylemma}}}
\\
&
& 
&
\\
&&& (i)\;\GI 
 \arrow[Leftarrow,bend right=8]{uur}[sloped,below]{\text{Prop.\,\ref*{prop:FSEtoGFE}}}
 \arrow[Leftarrow,bend left=8]{luu}[sloped,below]{\text{Prop.\,\ref*{prop:FSEtoGFE}}}
 \arrow[Rightarrow,bend right=15]{uurrr}[swap,sloped,below]{\text{Prop.\,\ref*{prop:GItoTRS}}}
\arrow[Rightarrow,bend left=17]{uulll}[swap,sloped,below]{\text{Prop.\,\ref*{prop:GItoTRS}}}
\end{tikzcd}
\caption{Proof of \cref{theo:TRSisos}}
\label{fig:TRSisos}   
\end{figure}

\begin{prop}[]
    \label{prop:FSEtoGFEinitiallylemma}
    $\Vd$-/$\Fd$-standard isomorphisms can be polynomially reduced to $\Vd$-/$\Fd$-global isomorphisms,
    $\SE\preceq^\Pc \GFE$ and $\VSE\preceq^\Pc \GVE$.
    \end{prop}
  \begin{proof}
    This follows immediately from \cref{lem:localisation} and polynomiality of $\Vd$-/$\Fd$-template calculation (\cref{lemma:template-computation}).
  \end{proof}

We employ the \emph{pointer-method} to reduce global TRS isomorphisms to graph isomorphisms.
$\Fd$/$\Vd$-labels in TRS-forests are encoded by isomorphically unique substructures, in this case a single vertex with properly chosen label.
Then every original vertex points towards the vertex representing its corresponding label and we use (strong) isomorphism on resulting OOLDGs.

\begin{prop}[]
\label{prop:FSEtoGFE}
$\Vd$-/$\Fd$-global isomorphisms can be polynomially reduced to the graph isomorphism problem,
$\GFE\preceq^\Pc \GI$ and $\GVE\preceq^\Pc \GI$.
\end{prop}
\begin{proof}
Let $R=(\Fd,\Vd,\Rd)$ be a TRS and $G(R)=(V,E,L,\lab)$ the corresponding TRS-forest
where $L=\Fd\cup \Vd\cup \{T\}\cup \N$. 
From $G(R)$ construct OOLDG $\Graph_\Fd(R)=(V',E',L',\lab')$ (or $\Graph_\Vd(R)=(V'',E'',L'',\lab'')$, resp.) as follows, where $L' = \Vd\cup \{F,T\}\cup \N_0$ (or $L'' = \Fd\cup \{C,T\}\cup \N_0$, resp.).
For $\Graph_\Fd(R)$, extend $V$ by new $F$-labelled vertices $v_f$ for each function symbol $f \in \Fd$, then for each $f$-labelled vertex add a new edge $(v,v_f,0)$ and modify the label of $v_f$ from $f$ to 0.
For $\Graph_\Vd(R)$ extend $V$ by new $C$-labelled vertices $v_x$ for each variable $x \in \Vd$,
then for each $x$-labelled vertex add a new edge $(v,v_x,0)$ and modify the label of $v_x$ from $x$ to 0.
For both constructions, $G(R)$ can be constructed  in polynomial time and the modifications can be done 
in at most $\Od(\sz_\Rd|\Rd|\log|\Fd|)$ or $\Od(\sz_\Rd|\Rd|\log|\Vd|)$ by 
traversing each tree breadth-first and checking whether a vertex is $\Fd$-/$\Vd$-labelled or not.

We will tackle $\GFE\preceq^\Pc \GI$ in detail.
 $\GVE\preceq^\Pc \GI$ is shown analogously, just switch the notions of symbol sets $\Fd$ and $\Vd$ and labels $F$ and $C$, 
and use $\Graph_\Vd(R)$ instead of $\Graph_\Fd(R)$.
By construction, $V'=V\cup \{v_f: f\in \Fd\}$, $E'=E\cup \{(v,v_f,0): v\in V, \lab(v){=}f{\in} \Fd\}$
and $\lab':V'\to \Vd\cup \{F,T\}\cup \N_0$ with
$\lab'
= \lab|_{V\setminus \lab^{-1}(\Fd)}
\opl \{v_f\mapsto F: f\in \Fd\} 
\opl \{v\mapsto 0 : v\in \lab^{-1}(\Fd)\}.$

Let $R_i=(\Fd_i,\Vd_i,\Rd_i)$, $i=1,2$, be TRSs, and let $G_i=\Graph_\Fd(R_i)$.
Now, $R_1\cong_{\SE}R_2$ iff $G_1\cong_S G_2$.
More precisely, $G_1\cong_S G_2$ is equivalent to $G(R_1)\cong G(R_2)$ \wrt~$\N\cup \Vd$-invariant map $\rho:L_1\to L_2$.
By \cref{cor:TRStreeiso} this is equivalent to $R_1\cong_{\GE}R_2$, witnessing a $\Vd$-invariant global isomorphism, but this is exactly $\cong_{\GFE}$.
Let $\phi': V_1'\to V_2'$ be a strong isomorphism between $G_1$ and $G_2$.
Then $|\Fd_1|{=}|\lab_1'^{-1}(\{F\})|{=}|\lab_2'^{-1}(\{F\})|=|\Fd_2|$, $|V_1|=|V_1'|-|\Fd_1|=|V_2'|-|\Fd_2|=|V_2|$, and thus $\rho:L_1\to L_2$, 
$\rho=\{f \mapsto f': f\in \Fd, v_{f'}=\phi'(v_f)\}\opl \id_{\{T\}\cup \Vd_1\cup \N}$
and $\phi=\phi'|_{V_1}:V_1\to V_2$, are well-defined and bijective maps.
Therefore $G(R_1)\cong G(R_2)$ via $\phi$ and \wrt~$\rho$ if we consider $\Fd$-labelled vertices to be unlabelled, and we just need to check the label property on this particular vertex subset.
Fix such an $\Fd$-labelled vertex $v\in V_1$, $\lab_1(v)=f\in \Fd_1$ with unique $0$-labelled outgoing edge $(v,v_f,0)\in E_1'$.
For $\phi(v)$, this corresponds to $(\phi(v),\phi'(v_f),0)\in E_2'$, but $\phi'(v_f)=v_{\rho(f)}$, which is already equivalent to $\rho(\lab_1(v))=\lab_2(\phi(v))$.

On the other hand, consider an isomorphism $\phi:V_1\to V_2$ \wrt~$\rho:L_1\to L_2$, where $\rho|_{\N\cup \Vd_1}=\id_{\N\cup \Vd_1}$ and in particular, $\rho(T)=T$ due to $\phi$ necessarily mapping roots onto roots.
Then $\rho|_{\Fd_1}:\Fd_1\to \Fd_2$ is bijective and can be used to extend $\phi$ to all of $V_1'$ via 
$\phi'=\phi \opl \{v_f\mapsto v_{\rho(f)}: f\in \Fd_1\}$.
Map $\phi'$ is well-defined and bijective, and due to construction, induces $G(R_1)\cong_S G(R_2)$, considered as appropriately relabelled subgraphs of $G_1$ and $G_2$ respectively.
Indeed, non-invariant $\Fd$-labels were purged in favour of uniform $0$-labels.
All $v_f$-vertices are $F$-labelled, so we just have to check isomorphism property on newly introduced $0$-labelled edges.
For $(v,v_f,0)\in E_1$, $\lab_1(v)=f$, and thus $(\phi(v),\phi'(v_f),0)\in E_2$, due to
$\phi'(v_f)=v_{\rho(f)}$ and $\lab_2(\phi(v))=\rho(\lab_1(v))$.
We are done, since by construction
$
|E_1'\setminus E_1|=\big|\lab_1'^{-1}(\{0\})\big|
=\big|\lab_1^{-1}(\Fd_1)\big|=\big|\lab_2^{-1}(\Fd_2)\big|
=\big|\lab_2'^{-1}(\{0\})\big|=\big|E_2'\setminus E_2\big|
$.
The claim now follows by \cref{prop:OOLDGScomplete}.
\end{proof}
\begin{figure}[htbp]
  \centering
  \begin{tikzpicture}[every node/.append style={thick,minimum size=0.7cm},xscale=0.6,yscale=0.8]
\node[shape=rectangle,draw=black] (v) at (1,-0.5)  {$T$};

\node[shape=circle,draw=black,fill=gray!20] (v111) at (-5,-4.5)  {$x$};
\node[shape=circle,draw=black,fill=gray!20] (v112) at (-3,-4.5)  {$y$};

\node[shape=circle,draw=black] (v11) at (-4,-3)  {$0$};
\node[shape=circle,draw=black,fill=gray!20] (v12) at (0,-3)  {$x$};

\node[shape=circle,draw=black] (v1) at (-2,-1.5)  {$0$};

\node[shape=circle,draw=black,fill=gray!20] (v21) at (2,-3)  {$x$};
\node[shape=circle,draw=black,fill=gray!20] (v22) at (6,-3)  {$y$};

\node[shape=circle,draw=black] (v2) at (4,-1.5)  {$0$};

\path [->,ultra thick,above] 
 (v1) edge node {$1$} (v11) 
 (v11) edge node[xshift=-1mm,yshift=-1mm] {$1$} (v111) 
 (v2) edge node[xshift=-2mm,yshift=-2mm] {$1$} (v21) 
;

\path [->,ultra thick,above]   
 (v2) edge node[xshift=2mm,yshift=-2mm] {$2$} (v22)
 (v1) edge node[xshift=2mm,yshift=-2mm] {$2$} (v12) 
 (v11) edge node[xshift=2mm,yshift=-1mm] {$2$} (v112)  
;

\path [->,ultra thick,above] (v) edge node {$2$} (v2);
\path [->,ultra thick,above] (v) edge node {$1$} (v1);

\node[shape=circle,draw=black,fill=gray!40] (vf) at (8.5,0.16)  {$F$};
\node[shape=circle,draw=black,fill=gray!40] (vg) at (8.5,-4)  {$F$};

%

\draw [->,ultra thick, rounded corners]  (v1.90) to ([yshift=12mm]v1.90) to node[above, pos=0.6,sloped] {$0$} (vf);

\draw [->, ultra thick,bend right=5]  (v11) to  node[below, pos=0.9,sloped] {$0$} (vg);
\draw [->, bend angle=20, bend left,ultra thick]  (v2) to node[above,pos=0.6,sloped] {$0$} (vg);

\node[shape=rectangle,draw=black] (w) at (16,-0.5)  {$T$};

\node[shape=circle,draw=black,fill=gray!20] (w11) at (11,-3)  {$y$};
\node[shape=circle,draw=black,fill=gray!20] (w12) at (15,-3)  {$x$};

\node[shape=circle,draw=black] (w1) at (13,-1.5)  {$0$};

\node[shape=circle,draw=black,fill=gray!20] (w21) at (18,-3)  {$x$};
\node[shape=circle,draw=black,fill=gray!20] (w22) at (20,-3)  {$y$};

\node[shape=circle,draw=black] (w2) at (19,-1.5)  {$0$};

\path [->,ultra thick,above] 
 (w1) edge node {$1$} (w11) 
 (w2) edge node[xshift=-2mm,yshift=-2mm] {$1$} (w21) 
;

\path [->,ultra thick,above]   
 (w2) edge node[xshift=2mm,yshift=-2mm] {$2$} (w22)
 (w1) edge node[xshift=2mm,yshift=-2mm] {$2$} (w12) 
;

\path [->,ultra thick,above] (w) edge node {$2$} (w2);
\path [->,ultra thick,above] (w) edge node {$1$} (w1);

\draw [->, bend angle=30, bend right,ultra thick]  (w1) to node[above,sloped,pos=0.6] {$0$} (vg);

\draw [->, ultra thick,rounded corners]  (w2.90) to ([yshift=12mm]w2.90)  to node[above,sloped,pos=0.6] {$0$} (vf);

\end{tikzpicture}
   \begin{tikzpicture}[every node/.append style={thick,minimum size=0.7cm},xscale=0.6,yscale=0.8]
\node[shape=rectangle,draw=black] (v) at (1,-0.5)  {$T$};

\node[shape=circle,draw=black,fill=gray!20] (v111) at (-5,-4.5)  {$0$};
\node[shape=circle,draw=black,fill=gray!20] (v112) at (-3,-4.5)  {$0$};

\node[shape=circle,draw=black] (v11) at (-4,-3)  {$h$};
\node[shape=circle,draw=black,fill=gray!20] (v12) at (0,-3)  {$0$};

\node[shape=circle,draw=black] (v1) at (-2,-1.5)  {$h$};

\node[shape=circle,draw=black,fill=gray!20] (v21) at (2,-3)  {$0$};
\node[shape=circle,draw=black,fill=gray!20] (v22) at (6,-3)  {$0$};

\node[shape=circle,draw=black] (v2) at (4,-1.5)  {$f$};

%
%
%
%
%
%

\path [->,ultra thick,above] 
 (v1) edge node {$1$} (v11) 
 (v2) edge node {$1$} (v21) 
;

\path [->,ultra thick,above]   
 (v2) edge node {$2$} (v22)
 (v1) edge node {$2$} (v12) 
;

\path [->,ultra thick,left] 
 (v11) edge node {$1$} (v111) 
;

\path [->,ultra thick,right] 
 (v11) edge node {$2$} (v112)  
;

\path [->,ultra thick,above] (v) edge node {$2$} (v2);
\path [->,ultra thick,above] (v) edge node {$1$} (v1);

\node[shape=circle,draw=black,fill=gray!40] (vx) at (4.5,-4.5)  {$C$};
\node[shape=circle,draw=black,fill=gray!40] (vy) at (12.5,-4.5)  {$C$};

\draw [->,rounded corners,ultra thick]  
(v12) |-
([xshift=-40mm]vx.center)
--
node[below] {} node[above] {$0$} 
([xshift=-20mm]vx.center)
--
(vx);
\draw [->, rounded corners,ultra thick]  
(v111) 
|- ([xshift=-80mm,yshift=-8mm]vx.center)
to node[below] {} node[above] {$0$} 
([xshift=-40mm,yshift=-8mm]vx.center)
-|
(vx);
\draw [->, rounded corners,ultra thick]  (v21) to node[below] {} node[above] {$0$} (vx);

\draw [->, rounded corners,ultra thick]  
(v112) 
|-
([xshift=-140mm,yshift=-12mm]vy.center)
to node[below] {} node[below] {$0$}
([xshift=-120mm,yshift=-12mm]vy.center)
-|
(vy);
\draw [->, rounded corners,ultra thick]  (v22) to node[below] {} node[above] {$0$} (vy);

\node[shape=rectangle,draw=black] (w) at (16,-0.5)  {$T$};

\node[shape=circle,draw=black,fill=gray!20] (w11) at (11,-3)  {$0$};
\node[shape=circle,draw=black,fill=gray!20] (w12) at (15,-3)  {$0$};

\node[shape=circle,draw=black] (w1) at (13,-1.5)  {$h$};

\node[shape=circle,draw=black,fill=gray!20] (w21) at (18,-3)  {$0$};
\node[shape=circle,draw=black,fill=gray!20] (w22) at (20,-3)  {$0$};

\node[shape=circle,draw=black] (w2) at (19,-1.5)  {$f$};

%
%
%
%
%

\path [->,ultra thick,above] 
 (w1) edge node {$1$} (w11) 
 (w2) edge node[xshift=-2mm,yshift=-2mm] {$1$} (w21) 
;

\path [->,ultra thick,above]   
 (w2) edge node[xshift=2mm,yshift=-2mm] {$2$} (w22)
 (w1) edge node[xshift=2mm,yshift=-2mm] {$2$} (w12) 
;

\path [->,ultra thick,above] (w) edge node {$2$} (w2);
\path [->,ultra thick,above] (w) edge node {$1$} (w1);

\draw [->, rounded corners,ultra thick]  
(w11) |- ([xshift=60mm]vx.center) to node[below] {} node[above] {$0$} (vx);

\draw [->,rounded corners,ultra thick]  
(w12) 
|-
([xshift=100mm,yshift=-17mm]vx.center)
to node[below] {} node[above] {$0$} 
([xshift=90mm,yshift=-17mm]vx.center)
--
([xshift=10mm,yshift=-17mm]vx.center)
--
(vx.-60);

\draw [->,rounded corners,ultra thick]  
(w21) 
|-
([xshift=120mm,yshift=-22mm]vx.center)
to node[below] {} node[above] {$0$}
([xshift=100mm,yshift=-22mm]vx.center)
--
([xshift=7mm,yshift=-22mm]vx.center)
--
(vx);

\draw [->,rounded corners,ultra thick]  (w22) 
|-
([xshift=70mm]vy.center)

to node[below] {} node[above,pos=0.1] {$0$} (vy);

\end{tikzpicture}
\caption{Example of encoding $\Graph_\Fd(R)$ (on top) and $\Graph_\Vd(R)$ (below) for TRS $\Rd=\{f(h(x,y),x)\to h(x,y), h(y,x)\to f(x,y)\}$ in the proof of  \cref{prop:FSEtoGFE}}
\label{fig:ex}
\end{figure}

\begin{prop}
\label{prop:GEtoGI}
Global isomorphisms can be polynomially reduced to the graph isomorphism problem,
$\GE\preceq^\Pc\GI$.
\end{prop}
\begin{proof}
Let $R=(\Fd,\Vd,\Rd)$ be a TRS.
Contrary to the $\Fd$/$\Vd$-global case in \cref{prop:FSEtoGFE}, we are not required to ensure invariance of one of the symbol sets.
According to \cref{cor:TRStreeiso}, we cannot outright use $G(R)$ and then apply an OOLDG isomorphism, even though this appears to be the easiest solution on paper.
A more nuanced encoding into an OOLDG $\Graph(R)=(V',E',L'=\N_0\cup \{z,g,F,C\},\lab')$ is thus to combine the encodings from \cref{prop:FSEtoGFE} into one:
construct the OOLDG $G(R) = (V,E,L,\lab)$ with $L=\Fd\cup \Vd\cup \{T\}\cup \N$,
add $C$-labelled vertices $v_x$ for each  $x \in \Vd$ and $F$-labelled vertices $v_f$ for each function symbol $f \in \Fd$.
For each vertex $v$, labelled with $x \in \Vd$, add an edge $(v,v_x,0)$  and change the label of $v$ from $x$ to $z$ (\ie~set~$\lab'(v)=z$).
For each vertex $v$ labelled with $f \in \Fd$, add an edge $(v,v_f,0)$ and change the label of $v$ from $f$ to fixed label $g$ (\ie~set~$\lab'(v) = g$).
This results in 
$V'=V\cup \{v_x: x\in \Vd\}\cup \{v_f: f\in \Fd\}$,
$E'=E\cup \big\{(v,v_x,0): v\in V, \lab(v)=x\in \Vd\big\}\cup \big\{(v,v_f,0): v\in V, \lab(v)=f\in \Fd\big\}$
and labelling function $\lab':V'\to \N_0\cup \{z,g,F,C\}$ with
\begin{align*}
\lab'
& = \lab|_{\lab^{-1}(\{T\})}
\opl\{v_x\mapsto C: x\in \Vd\}\opl \{v_f\mapsto F: f\in \Fd\}\\
& \quad \;\opl \{v\mapsto z: v\in \lab^{-1}(\Vd)\}
\opl \{v\mapsto g: v\in \lab^{-1}(\Fd)\}.
\end{align*}
The construction can be  done in polynomial time by computing $G(R)$,
performing the modifications by traversing each tree breadth-first,
and checking whether a vertex is $\Vd$-/$\Fd$-labelled.   
Let $G_i=\Graph(R_i)$, for two TRSs $R_i=(\Fd_i,\Vd_i,\Rd_i)$, $i=1,2$.
Then $G_1\cong_S G_2$ iff $G(R_1)\cong G(R_2)$, \wrt~$\N$-invariant and symbol-relationship preserving $\rho:L_1\to L_2$ (that is $\rho(\Vd_1)=\Vd_2$ and 
$\rho(\Fd_1)=\Fd_2$), 
and is a direct consequence of \cref{cor:TRStreeiso}.
We accomplish label-type differentiation by introducing two distinct labels to encode additionally introduced vertices.
Relabelling of original vertices, also with distinct labels, simplifies the proof.
First, fix a strong isomorphism $\phi': V_1'\to V_2'$.
Then $\rho: L_1\to L_2$ with 
\begin{equation*}
\rho=\id_{\N\cup \{T\}}\opl\sset{f\mapsto f': f\in \Fd_1,\phi'(v_f)=v_{f'}}\opl \sset{x\mapsto x': x\in \Vd_1,\phi'(v_x)=v_{x'}}
\end{equation*}
 is well-defined and bijective
due to label-preserving property, \ie~$|\{v_f: f\in \Fd_1\}|=|\lab_1'^{-1}(\{F\})|=|\lab_2'^{-1}(\{F\})|=|\{v_f: f\in \Fd_2\}|$, analogously for vertices $v_x$, where $x$ is an arbitrary variable symbol in either $\Vd_1$ or $\Vd_2$.
Moreover, $V_i=\lab_i'^{-1}(\{z,g\})$, $i=1,2$, and  $\phi=\phi'|_{V_1}:V_1\to V_2$ is  an isomorphism \wrt~$\rho$.
In this case, $\phi$ and $\rho$ already satisfy all assumptions of the reverse implication in \cref{cor:TRStreeiso}, since 
$\Vd_2=\rho(\Vd_1)$ and $\lab_1(v)\in \Vd_1$ iff $\lab_1'(v)=z$.
\begin{figure}[t] 
  \centering
\newcommand{\HLABEL}{g}
\newcommand{\ZLABEL}{z}
  \begin{tikzpicture}[every node/.append style={thick,minimum size=0.7cm},xscale=0.60,yscale=0.8]
\node[shape=rectangle,draw=black] (v) at (1,-0.5)  {$T$};

\node[shape=circle,draw=black,fill=gray!20] (v111) at (-5,-4.5)  {$\ZLABEL$};
\node[shape=circle,draw=black,fill=gray!20] (v112) at (-3,-4.5)  {$\ZLABEL$};

\node[shape=circle,draw=black] (v11) at (-4,-3)  {$\HLABEL$};
\node[shape=circle,draw=black,fill=gray!20] (v12) at (0,-3)  {$\ZLABEL$};

\node[shape=circle,draw=black] (v1) at (-2,-1.5)  {$\HLABEL$};

\node[shape=circle,draw=black,fill=gray!20] (v21) at (2,-3)  {$\ZLABEL$};
\node[shape=circle,draw=black,fill=gray!20] (v22) at (6,-3)  {$\ZLABEL$};

\node[shape=circle,draw=black] (v2) at (4,-1.5)  {$\HLABEL$};

%
%
%
%
%
%

\path [->,ultra thick,above] 
 (v1) edge node {$1$} (v11) 
 (v2) edge node {$1$} (v21) 
;

\path [->,ultra thick,above]   
 (v2) edge node {$2$} (v22)
 (v1) edge node {$2$} (v12) 
;

\path [->,ultra thick,left] 
 (v11) edge node {$1$} (v111) 
;

\path [->,ultra thick,right] 
 (v11) edge node {$2$} (v112)  
;

\path [->,ultra thick,above] (v) edge node {$2$} (v2);
\path [->,ultra thick,above] (v) edge node {$1$} (v1);

\node[shape=circle,draw=black,fill=gray!40] (vf) at (4.5,0.7)  {$F$};
\node[shape=circle,draw=black,fill=gray!40] (vg) at (13,0)  {$F$};

\node[shape=circle,draw=black,fill=gray!40] (vx) at (4.5,-4.5)  {$C$};
\node[shape=circle,draw=black,fill=gray!40] (vy) at (12.5,-4.5)  {$C$};

\draw [->,rounded corners,ultra thick]  (v12)
|- ([xshift=-24mm]vx.center) node[below] {} node[below] {$0$} --  (vx);
\draw [->, rounded corners,ultra thick]  (v111) |- ([yshift=-8mm,xshift=-70mm]vx.center) to node[below] {} node[below] {$0$} ([yshift=-8mm,xshift=-30mm]vx.center)
--
([yshift=-8mm,xshift=-10mm]vx.center)
-- (vx.-140);
\draw [->, ultra thick]  (v21) to node[below] {} node[above] {$0$} (vx);

\draw [->,  rounded corners,ultra thick]  (v112) |- ([yshift=-24mm,xshift=-140mm]vy.center) -- 
node[below] {} node[above] {$0$} 
([yshift=-24mm,xshift=-40mm]vy.center) 
-|
(vy);
\draw [->, rounded corners,ultra thick]  (v22) to node[below] {} node[above] {$0$} (vy);

\draw [->,ultra thick, rounded corners]  (v1) |-
([xshift=-50mm]vf.center)
-- node[below] {$0$} 
 ([xshift=-20mm]vf.center)
--
(vf);
\draw [->,  ultra thick, rounded corners]  (v2) |-
([xshift=-30mm]vg.center)
to node[above, pos=0.6] {$0$} (vg);
\draw [->,ultra thick, rounded corners]  (v11)
|-  ([xshift=-166mm,yshift=15mm]vg.center) to node[below, pos=0.6] {$0$} 
([xshift=-146mm,yshift=15mm]vg.center)
-|
(vg);

\node[shape=rectangle,draw=black] (w) at (16,-0.5)  {$T$};

\node[shape=circle,draw=black,fill=gray!20] (w11) at (11,-3)  {$\ZLABEL$};
\node[shape=circle,draw=black,fill=gray!20] (w12) at (15,-3)  {$\ZLABEL$};

\node[shape=circle,draw=black] (w1) at (13,-1.5)  {$\HLABEL$};

\node[shape=circle,draw=black,fill=gray!20] (w21) at (18,-3)  {$\ZLABEL$};
\node[shape=circle,draw=black,fill=gray!20] (w22) at (20,-3)  {$\ZLABEL$};

\node[shape=circle,draw=black] (w2) at (19,-1.5)  {$\HLABEL$};

%
%
%
%
%

\path [->,ultra thick,above] 
 (w1) edge node {$1$} (w11) 
 (w2) edge node[xshift=-2mm,yshift=-2mm] {$1$} (w21) 
;

\path [->,ultra thick,above]   
 (w2) edge node[xshift=2mm,yshift=-2mm] {$2$} (w22)
 (w1) edge node[xshift=2mm,yshift=-2mm] {$2$} (w12) 
;

\path [->,ultra thick,above] (w) edge node {$2$} (w2);
\path [->,ultra thick,above] (w) edge node {$1$} (w1);

\draw [->,  rounded corners,ultra thick]  (w11) |- ([xshift=60mm]vx.center) to node[below] {} node[below] {$0$} (vx);

\draw [->, rounded corners,ultra thick]  (w12) |- ([yshift=-8mm,xshift=70mm]vx.center) -- node[below] {} node[below] {$0$} 
([yshift=-8mm,xshift=30mm]vx.center)
-- ([yshift=-8mm,xshift=10mm]vx.center)
-- (vx.-60);

\draw [->, rounded corners, ultra thick]  (w21) |- ([yshift=-16mm,xshift=70mm]vx.center) node[below] {} node[below] {$0$} 
-- ([yshift=-16mm,xshift=30mm]vx.center) -| (vx);

\draw [->,  rounded corners,ultra thick]  (w22) |- 
([xshift=70mm]vy.center) --
node[below] {} node[below] {$0$} 
([xshift=34mm]vy.center) 
-- (vy);


\draw [->, ultra thick,rounded corners]  (w2) |- ([xshift=140mm,yshift=12mm]vf.center) -- node[below,pos=0.6] {$0$} 
([xshift=80mm,yshift=12mm]vf.center)
-|
(vf);
\draw [->, ultra thick,rounded corners]  (w1.90) to (w1)  to node[left,pos=0.6] {$0$} (vg);

\end{tikzpicture}

\caption{$\Graph(R)$ for TRS $\Rd=\{f(h(x,y),x)\to h(x,y), h(y,x)\to f(x,y)\}$ in the proof of  \cref{prop:GEtoGI}}
\label{fig:test2}
\end{figure}  
By vertex set restriction, $G(R_1)\cong_S G(R_2)$
if we consider vertices to be unlabelled, with notable exception of $T$-labelled roots.
Fix $\Fd$-labelled vertex $v\in V_1$, $\lab_1(v)=f$ with unique $0$-labelled outgoing edge $(v,v_f,0)\in E_1'$.
Just as in \cref{prop:FSEtoGFE}, $\rho(\lab_1(v))=\lab_2(\phi(v))$ by inspecting $0$-labelled outgoing edges of $\phi(v)$.
The case of $\Vd$-labelled vertices is analogous, 
distinguishing the two cases by different labels $z$ and $g$ in $G(R_i)$.
Consequently, $G(R_1)\cong G(R_2)$ via~$\phi$ and \wrt~$\rho$.

On the contrary,
let $\phi: V_1\to V_2$ be an isomorphism \wrt~$\rho:L_1\to L_2$, where $\rho|_{\N}=\id_\N$ and $\rho(\Vd_1)=\Vd_2$.
Then $\rho(T)=T$, since $\phi$ maps roots to roots.
Extend $\phi$ to all of $V_1'$ via 
$\phi' = \phi \opl \sset{v_f\mapsto v_{\rho(f)}: f\in \Fd_1} \opl \sset{v_x\mapsto v_{\rho(x)}: x\in \Vd_1}$,
well-defined and bijective due to $|\Fd_1|=|\Fd_2|$ and $|\Vd_1|=|\Vd_2|$.
This again induces $G(R_1)\cong_SG(R_2)$, understood as appropriately relabelled subgraphs of $G_1$ and $G_2$.
Indeed, if $\lab_1(v)\in \Fd_1$,
$\lab_2(\phi(v))=\rho(\lab_1(v))\in \Fd_2$ and hence $\lab_1'(v)=g=\lab_2'(\phi(v))$, analogously for $\lab_1(v)\in \Vd_1$.
It is left to check isomorphism property on newly introduced $0$-labelled edges.
For fix non-root vertex $v\in V_1$, this is proven the same way as in \cref{prop:FSEtoGFE}, where again we can distinguish the two cases $\Fd$/$\Vd$-labelled by different labels $z$ and $g$ in $G(R_i)$.
The claim then follows by \cref{prop:OOLDGScomplete}.
\end{proof}


\begin{prop}[]
\label{prop:GItoTRS}
$\GI\preceq^\Pc \VSE$, $\GI\preceq^\Pc\SE$ and $\GI\preceq^\Pc\GE$.
\end{prop}
\begin{proof}
By \cref{prop:OOLDGScomplete}, it suffices to consider an unlabelled directed graph $G=(V,E)$.
Let  $R(G)=(\Fd,\Vd,\Rd)$ and $R'(G)=(\Fd',\Vd',\Rd')$ be encodings with:
\begin{itemize}
  \itemsep0em
 \item[(1)] $\Fd=\{f_v: v\in V\}\cup \{c\}$, $\Vd=\{x\}$, $\Rd=\{f_v(f_w(x))\to c: (v,w)\in E\}$ and
 \item[(2)] $\Fd'=\{f,c\}$, $\Vd'=\{x_v: v\in V\}$, $\Rd'=\{f(x_v,x_w)\to c: (v,w)\in E\}$.
\end{itemize}
Note that $f,c,x$ are fixed for respective encoding, while function symbols $f_v$ and variables $x_v$ depend on vertices $v\in V$.
Vertices are represented by either unique function symbols or variables, while edges are encoded by rewriting rules.
See \cref{fig:DGtoTRS} for an example.
Two unlabelled directed graphs $G_1$ and $G_2$ are isomorphic iff 
$R_1\cong_{\GE}R_2$, or iff $R_1'\cong_{\VSE}R_2'$, where $R_i=R(G_i)$ and $R_i'=R'(G_i)$.
Moreover, due to construction, $R_1\cong_{\GE} R_2$ is equivalent to $R_1\cong_{\SE} R_2$.
\begin{figure}[t]
\centering
\begin{subfigure}{.5\textwidth}
  \centering
  \begin{tikzpicture}[every node/.append style={thick},scale=0.5]
\node[shape=circle,draw=black] (v1) at (0,0)  {$v_1$};
\node[shape=circle,draw=black] (v2) at (12,0)  {$v_2$};
\node[shape=circle,draw=black] (v3) at (12,-6) {$v_3$};
\node[shape=circle,draw=black] (v4) at (0,-6) {$v_4$};
\node[shape=circle,draw=black] (v5) at (6,-3) {$v_5$};

\path [->,ultra thick] 
 (v1) edge node {} (v2) 
 (v2) edge node {} (v3) 
 (v3) edge node {} (v4) 
 (v4) edge node {} (v1) 
 (v1) edge node {} (v5) 
 (v5) edge node {} (v2) 
;

\end{tikzpicture}
  \caption{Directed graph $G$ of order $5$ and size $6$}
  \label{fig:DGtoTRSb}
\end{subfigure}%
\hfill
\begin{subfigure}{.5\textwidth}
  \centering$
  \begin{array}{@{}l@{\,}ll}

  \Fd = \{ & f_1,f_2,f_3,f_4,f_5,c\}\\
  \Rd = \{ 
  & f_{1}(f_{2}(x))\to c, f_{1}(f_{5}(x))\to c,\\
  & f_{2}(f_{3}(x))\to c, f_{3}(f_{4}(x))\to c,\\
  & f_{4}(f_{1}(x))\to c, f_{5}(f_{2}(x))\to c\}
  \end{array}$
  \\
  $\begin{array}{l@{\,}ll}

\Vd' = \{ & x_1,x_2,x_3,x_4,x_5\}\\
  \Rd' = \{
  & f(x_{1},x_{2})\to c,f(x_{1},x_{5})\to c,\\
  & f(x_{2},x_{3})\to c,f(x_{3},x_{4})\to c,\\
  & f(x_{4},x_{1})\to c,f(x_{5},x_{2})\to c\}
  \end{array}$
  \caption{Encodings $R(G)$ and $R'(G)$}
  \label{fig:DGtoTRSa}
\end{subfigure}
\caption{$R(G)$ and $R'(G)$ in the proof of  \cref{prop:GItoTRS}, where function symbol $f_i$ and variable $x_i$, resp., refer to vertex $v_i$
\label{fig:DGtoTRS}}
\end{figure}
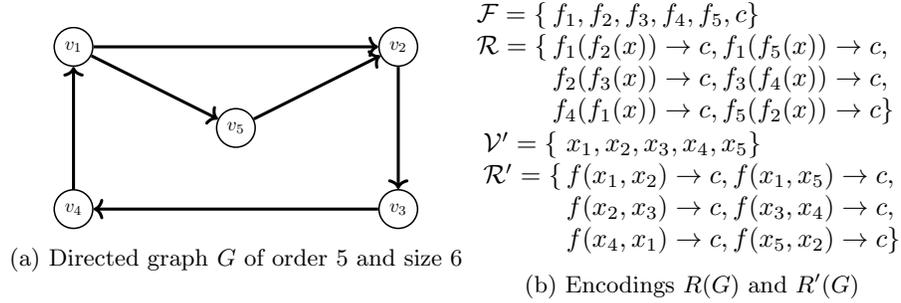

{W.l.o.g.}~we can assume that there exist no vertices with no incoming and no outgoing edges.
If $\phi: V_1\to V_2$ is a graph isomorphism between $G_1,G_2$, one easily checks that 
$\phi':\Fd_1\to \Fd_2$, $\phi'(x)=x$, $\phi'(c)=c$ and $\phi'(f_v)=f_{\phi(v)}$ is a well-defined global isomorphism, and
$\psi':(\Fd'\cup \Vd_1')\to (\Fd'\cup \Vd_2')$, $\psi'(x_v)=x_{\phi(v)}$ a well-defined $\Vd$-standard isomorphism.

For the reverse implication,
first assume $\phi':\Fd_1\to \Fd_2$ to be a global isomorphism between $R_1$ and $R_2$.
Since $\phi'$ preserves arity of function symbols, 
map $\phi:V_1\to V_2$, so that $\phi'(f_v)=f_{\phi(v)}$, is well-defined and bijective due to $|V_1|=|\Fd_1\setminus \{c\}|=|\Fd_2\setminus \{c\}|=|V_2|$ and $\phi'$ invariant on $\{x,c\}$.
Fix $(v,w)\in E_1$.
Then $f_{\phi(v)}(f_{\phi(w)}(x))\to c = \phi'(f_v)(\phi'(f_w)(x))\to c\in \Rd_2$,
which is equivalent to $(\phi(v),\phi(w))\in E_2$.
Since 
$|E_1|=|\Rd_1|=|\Rd_2|=|E_2|$, we are done.

Now assume $\psi':(\Fd'\cup \Vd_1')\to (\Fd'\cup \Vd_2')$ is a $\Vd$-standard isomorphism, or equivalently, a $\Vd$-global isomorphism (singleton variable set), between $R_1'$ and $R_2'$.
Again, since $\psi'$ preserves the arity, we have $\psi'(f)=f$ and $\psi'(c)=c$.
Thus map $\phi:V_1\to V_2$, so that $\psi'(x_v)=x_{\phi(v)}$ is well-defined and bijective due to 
$|V_1|=|\Vd_1'|=|\Vd_2'|=|V_2|$.
Fix $(v,w)\in E$.
Then analogously to the case before, $(\phi(v),\phi(w))\in E_2$ and since $|E_1|=|\Rd_1'|=|\Rd_2'|=|E_2|$, we are done.
\end{proof}

\begin{proof}[Proof of \cref{theo:TRSisos}]
The polynomial reductions as depicted in \cref{fig:TRSisos} follow immediately by
 Propositions~\ref{prop:FSEtoGFEinitiallylemma}, \ref{prop:FSEtoGFE}, \ref{prop:GEtoGI}, and \ref{prop:GItoTRS}.
\end{proof}

\begin{remark}
Generalised TRS isomorphisms remain in the same complexity class as their ungeneralised counterparts.
In particular, determining generalised-standard TRS isomorphisms is also $\GI$-complete.
Moreover, 
the complexity classes of generalised and ungeneralised TRS isomorphisms also do not change when we drop requirements on rewriting rules in accordance to \cref{rem:rsubseteql}.
\end{remark}

\section{Conclusion\label{sect:concl}}
We considered eight notions of syntactic equality up to {\renaming} of function symbols and/or variables 
for TRSs. We have shown for each of them that they are either efficiently decidable or they are $\GI$-complete. 
This clarifies the complexity of the equalities.
We discussed semantic properties of the syntactic equivalences briefly. The implications of this study are that for an adequate notion, renamings of functions should be global, while renaming of variables can be done locally, globally or not at all.

Instead of explicitly stating encodings for every global TRS isomorphism-case, we took advantage of the general structure of TRSs in form of templates and, in particular, the relationship between variables and function symbols. Presented template-generating-algorithms can easily be generalised to handle more than two disjoint symbol sets and can then be applied for reductions or polynomial solvability proofs in the context of new isomorphisms if they demand symbol-relationship to be preserved. In fact, we saw that under canonical remapping of symbols, proof of local TRS isomorphisms reduces to simple set comparison.


\bibliography{references}

\begin{thebibliography}{10}
\expandafter\ifx\csname url\endcsname\relax
  \def\url#1{\texttt{#1}}\fi
\expandafter\ifx\csname urlprefix\endcsname\relax\def\urlprefix{URL }\fi
\expandafter\ifx\csname href\endcsname\relax
  \def\href#1#2{#2} \def\path#1{#1}\fi

\bibitem{Brockschmidt-et-al:2021}
M.~Brockschmidt, R.~Musiol, C.~Otto, J.~Giesl, Automated termination proofs for
  java programs with cyclic data, in: P.~Madhusudan, S.~A. Seshia (Eds.),
  Computer Aided Verification - 24th International Conference, {CAV} 2012,
  Berkeley, CA, USA, July 7-13, 2012 Proceedings, Vol. 7358 of Lecture Notes in
  Computer Science, Springer, 2012, pp. 105--122.
\newblock \href {https://doi.org/10.1007/978-3-642-31424-7\_13}
  {\path{doi:10.1007/978-3-642-31424-7\_13}}.

\bibitem{Giesl-et-al:2011}
J.~Giesl, M.~Raffelsieper, P.~Schneider{-}Kamp, S.~Swiderski, R.~Thiemann,
  Automated termination proofs for haskell by term rewriting, {ACM} Trans.
  Program. Lang. Syst. 33~(2) (2011) 7:1--7:39.
\newblock \href {https://doi.org/10.1145/1890028.1890030}
  {\path{doi:10.1145/1890028.1890030}}.

\bibitem{Schoening:88}
U.~Sch{\"{o}}ning, Graph isomorphism is in the low hierarchy, J. Comput. Syst.
  Sci. 37~(3) (1988) 312--323.
\newblock \href {https://doi.org/10.1016/0022-0000(88)90010-4}
  {\path{doi:10.1016/0022-0000(88)90010-4}}.

\bibitem{Kobler-et-al:92}
J.~K{\"{o}}bler, U.~Sch{\"{o}}ning, J.~Tor{\'{a}}n, Graph isomorphism is low
  for {PP}, Comput. Complex. 2 (1992) 301--330.
\newblock \href {https://doi.org/10.1007/BF01200427}
  {\path{doi:10.1007/BF01200427}}.

\bibitem{grohe-schweitzer:2020}
M.~Grohe, P.~Schweitzer, The graph isomorphism problem, Commun. ACM 63~(11)
  (2020) 128–134.
\newblock \href {https://doi.org/10.1145/3372123} {\path{doi:10.1145/3372123}}.

\bibitem{booth-colbourn:77}
K.~S. Booth, C.~J. Colbourn,
  \href{https://cs.uwaterloo.ca/research/tr/1977/CS-77-04.pdf}{Problems
  polynomially equivalent to graph isomorphism}, Tech. Rep. CS-77-04,
  University of Waterloo (1979).
\newline\urlprefix\url{https://cs.uwaterloo.ca/research/tr/1977/CS-77-04.pdf}

\bibitem{zemlyachenko-et-al:85}
V.~N. Zemlyachenko, N.~M. Korneenko, R.~I. Tyshkevich, Graph isomorphism
  problem, J. Math. Sci. (N. Y.) 29 (1985) 1426--1481.
\newblock \href {https://doi.org/10.1007/BF02104746}
  {\path{doi:10.1007/BF02104746}}.

\bibitem{Toyama86}
Y.~Toyama, How to prove equivalence of term rewriting systems without
  induction, in: J.~H. Siekmann (Ed.), 8th International Conference on
  Automated Deduction, Oxford, England, July 27 - August 1, 1986, Proceedings,
  Vol. 230 of Lecture Notes in Computer Science, Springer, 1986, pp. 118--127.
\newblock \href {https://doi.org/10.1007/3-540-16780-3\_84}
  {\path{doi:10.1007/3-540-16780-3\_84}}.

\bibitem{plaisted:93}
D.~A. Plaisted, Equational reasoning and term rewriting systems, in: Handbook
  of Logic in Artificial Intelligence and Logic Programming (Vol. 1), Oxford
  University Press, USA, 1993, Ch.~5, pp. 274--364.

\bibitem{Terese03}
Terese, Term Rewriting Systems, Vol.~55 of Cambridge Tracts in Theoretical
  Computer Science, Cambridge University Press, 2003.

\bibitem{Hirokawa-Middeldorp-Sternagel:14}
N.~Hirokawa, A.~Middeldorp, C.~Sternagel,
  \href{http://www.nue.ie.niigata-u.ac.jp/toyama/iwc2014/iwc2014.pdf}{Normalization
  equivalence of rewrite systems}, in: T.~Aoto, D.~Kesner (Eds.), Proceedings
  IWC 2014, 3rd International Workshop on Confluence, 2014, pp. 14--18.
\newline\urlprefix\url{http://www.nue.ie.niigata-u.ac.jp/toyama/iwc2014/iwc2014.pdf}

\bibitem{Metivier83}
Y.~M{\'{e}}tivier, About the rewriting systems produced by the knuth-bendix
  completion algorithm, Inf. Process. Lett. 16~(1) (1983) 31--34.
\newblock \href {https://doi.org/10.1016/0020-0190(83)90009-1}
  {\path{doi:10.1016/0020-0190(83)90009-1}}.

\bibitem{Mohan90}
C.~K. Mohan, Equivalences of rewrite programs, in: S.~Kaplan, M.~Okada (Eds.),
  Conditional and Typed Rewriting Systems, 2nd International {CTRS} Workshop,
  Montreal, Canada, June 11-14, 1990, Proceedings, Vol. 516 of Lecture Notes in
  Computer Science, Springer, 1990, pp. 92--97.
\newblock \href {https://doi.org/10.1007/3-540-54317-1\_82}
  {\path{doi:10.1007/3-540-54317-1\_82}}.

\bibitem{basin:1994}
D.~A. Basin, A term equality problem equivalent to graph isomorphism,
  Information Processing Letters 51~(2) (1994) 61--66.
\newblock \href {https://doi.org/https://doi.org/10.1016/0020-0190(94)00084-0}
  {\path{doi:https://doi.org/10.1016/0020-0190(94)00084-0}}.

\bibitem{schmidt-schauss-et-al:2013}
M.~Schmidt{-}Schau{\ss}, C.~Rau, D.~Sabel, Algorithms for extended
  alpha-equivalence and complexity, in: F.~van Raamsdonk (Ed.), 24th
  International Conference on Rewriting Techniques and Applications, {RTA}
  2013, June 24-26, 2013, Eindhoven, The Netherlands, Vol.~21 of LIPIcs,
  Schloss Dagstuhl - Leibniz-Zentrum f{\"{u}}r Informatik, 2013, pp. 255--270.
\newblock \href {https://doi.org/10.4230/LIPIcs.RTA.2013.255}
  {\path{doi:10.4230/LIPIcs.RTA.2013.255}}.

\bibitem{rosenkrantz-hunt:85}
D.~J. Rosenkrantz, H.~B.~H. III, Testing for grammatical coverings, Theor.
  Comput. Sci. 38 (1985) 323--341.
\newblock \href {https://doi.org/10.1016/0304-3975(85)90226-9}
  {\path{doi:10.1016/0304-3975(85)90226-9}}.

\bibitem{baader-nipkow:98}
F.~Baader, T.~Nipkow, Term rewriting and all that, Cambridge University Press,
  1998.

\bibitem{ohlebusch:2002}
E.~Ohlebusch, Advanced topics in term rewriting, Springer, 2002.

\end{thebibliography}

\end{document}